\documentclass[twocolumn,aps,prb,showpacs,a4paper,superscriptaddress,longbibliography]{revtex4-2}

\usepackage{graphicx}
\usepackage{multirow}
\usepackage[dvipsnames]{xcolor}
\usepackage{amsmath,amsfonts,amssymb}
\usepackage{xr-hyper}
\usepackage[hidelinks]{hyperref}

\begin{document}

\title{The rich structural, electronic and bonding landscape of 1\texorpdfstring{$T$}~-type \texorpdfstring{TaTe$_2$}~ single-layers}

\author{Jose Angel Silva-Guill\'{e}n}
\email{joseangel.silva@imdea.org} 
\affiliation{Fundaci\'{o}n IMDEA Nanociencia, C/ Faraday 9, Campus Cantoblanco, 28409 Madrid, Spain}

\author{Enric Canadell}
\affiliation{Institut de Ci\`encia de Materials de Barcelona (ICMAB-CSIC), Campus UAB, 08193 Bellaterra, Spain, and, Royal Academy of Sciences and Arts of Barcelona, Chemistry Section, La Rambla 115, 08002 Barcelona, Spain}

\begin{abstract}
Charge density waves (CDW) in single-layer 1$T$-MTe$_2$ (M= Nb, Ta) recently raised large attention because of the contrasting structural and physical behavior with the sulfide and selenide analogues. A first-principles study of fourteen different 1$T$-type TaTe$_2$ single-layers is reported. The importance of Te to Ta electron transfer and multicenter metal-metal bonding in stabilizing different structural modulations is highlighted. Analysis of the electronic structure of the optimized structures provides a rationale for what distinguishes 1$T$-TaTe$_2$ from the related disulfide and diselenide, what are the more stable structural modulations for 1$T$-type TaTe$_2$ single-layers, the possible role of Fermi surface nesting on some of these CDW instabilities, how the CDW affects the metallic properties of the non-distorted lattice and the possibility that some of these CDW phases exhibit exotic properties. All CDW phases studied
exhibit band structures typical of metallic systems although some exhibit both very flat and dispersive bands at the Fermi level so that Mott effects could develop; one of the (4$\times$4) phases studied exhibits a Dirac cone at the Fermi level.

\end{abstract}


\maketitle

\section{Introduction}

The field of charge density wave (CDW) materials has experienced a spectacular development in the last decade. Both new materials and challenging properties have been reported~\cite{Pouget2024,Balandin2021,Monceau2012,Pouget2016}.
Importantly, some of these properties have even paved the way for practical applications~\cite{Balandin2022,Xu2021}. Group V transition metal dichalcogenides (TMDC) have played a central part in our understanding of the diversity of electronic instabilities in low-dimensional conductors. The discovery of the structural modulations in bulk groups IV and V TMDC almost fifty years ago~\cite{Wilson1975,Scruby1975} played a decisive role in calling attention to the rich and novel physics of very anisotropic conductors. Group V TMDC are built from MX$_2$ layers (M= V, Nb, Ta; X= S, Se, Te) where M can exhibit either an octahedral (1$T$-type polymorphs) or trigonal prismatic coordination (2$H$-type polymorphs). These systems have been and remain controversial for many reasons~\cite{Rossnagel2011}. For instance, the possibility of either strong or weak electron–phonon coupling CDW scenarios~\cite{Pouget2024} and the competition between CDW and superconductivity in 2$H$-MX$_2$ (M = Nb, Ta; X = S, Se) have been discussed for decades. The MX$_2$ layers interact through weak van der Waals forces and thus they are easily exfoliated. Consequently, it is possible to examine the more interesting and controversial issues at the two-dimensional (2D) limit as well as by smoothly changing the density of carriers through gate doping. This is at the origin of the impressive revival of interest recently raised by these materials~\cite{Chhowalla2013,Manzeli2017,Roldan_ChSR_2017}. Indeed, intriguing differences of these few-layers or even single-layer materials with their bulk counterparts have been discovered~\cite{Li2016,Si2020}.

An interesting aspect which is now the object of large attention is the difference between group V MX$_2$ tellurides from sulfides and selenides. In general, the valence bands of transition metal tellurides are noticeably wider than those of selenides or sulfides because the Te 5$p$ orbitals are considerably more diffuse than the S 3$p$ and Se 4$p$ ones. Therefore, there is often a substantial overlap of the Te-based valence bands with the bottom part of the transition metal-based $d$ bands leading to a non-negligible electron transfer from tellurium to the transition metal atoms~\cite{Canadell1992,Whangbo1992}. These electron transfers have strong implications for the structural and transport properties of many transition metal tellurides which often exhibit crystal structures and transport properties differing from those of the corresponding selenides or sulphides~\cite{Meerschaut1986}. Bulk MX$_2$ (M= Nb, Ta; X= S, Se) are found in both the 1$T$ and 2$H$ hexagonal polymorphs and in both cases they exhibit CDW. In contrast, bulk MTe$_2$ (M= V, Nb, Ta) are only found in a strongly distorted monoclinic 1$T$'-type structure~\cite{Brown1966}. From now on except when otherwise stated we will always refer to 1$T$-type (octahedral) TaTe$_2$ structures.

\begin{figure*}[t]
    \centering
   \includegraphics{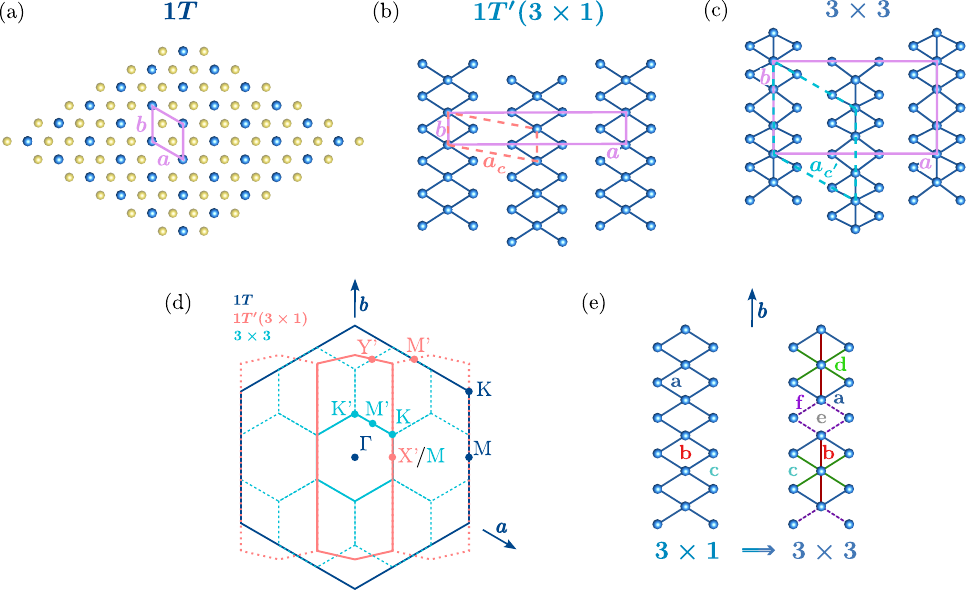}
   \caption{Top layer of the bulk structure of 1$T$-TaTe$_2$ (a), 1$T$'-TaTe$_2$ (i.e. (3$\times$1)-TaTe$_2$) (b) and (3$\times$3)-TaTe$_2$ (c). For simplicity, the Te atoms are not shown in (b) and (c). The dashed red (b) and blue (c) repeat units are those used for the (3$\times$1) and (3$\times$3) single-layer calculations. In (d) the relationship between the BZ of the three single-layers is schematically shown. The different Ta-Ta bonds referred to in the discussion are shown in (e).}
    \label{fig:struct_1}
\end{figure*}

Bulk TaTe$_2$ at room temperature is metallic~\cite{Vernes1998,Chen2017,Li2018} and exhibits a distorted 1$T$’ structure~\cite{Brown1966} containing very peculiar double zigzag chains in the Ta sublattice (see Fig.~\ref{fig:struct_1}b). Such a distinctive structure, only shown by group V MTe$_2$ (M= V, Nb, Ta), is thought to occur as a result of the above mentioned large Te to Ta electron transfer~\cite{Whangbo1992}. Because of the trimeric zigzag chains, each layer of 1$T$’-TaTe$_2$ can be thought of as a (3$\times$1) reconstruction of an ideal, non distorted 1$T$-TaTe$_2$ structure (Fig.~\ref{fig:struct_1}a). Bulk 1$T$’-TaTe$_2$ undergoes a structural modulation around 170 K below which the zigzag trimeric chains rearrange and the periodicity along the chains becomes three times larger~\cite{Sorgel2006}. Consequently, below 170 K the layers can be described as a (3$\times$3) reconstruction of 1$T$-TaTe$_2$ (Fig.~\ref{fig:struct_1}c). Let us note that the
emergence of a ($\sqrt19 \times \sqrt19$) periodicity in bulk has already been known for some time~\cite{Landuyt1974,Landuyt1978} although a full structural characterization of the latter was not reported. The metallic character of 1$T$'-TaTe$_2$ is kept across the 170 K transition, although anomalies in the resistivity, magnetic susceptibility and specific heat capacity have been detected~\cite{Sorgel2006,Liu2015,Chen2017}. It is worth noting that bulk 1$T$'-NbTe$_2$ does not exhibit the low-temperature transition towards the  (3$\times$3) modulated structure. 1$T$'-NbTe$_2$ is an ambient pressure superconductor ($T_c$= 0.4 K)~\cite{Nagata1993} but (3$\times$3)-TaTe$_2$ it is not. However, under pressure, when apparently the (3$\times$3) CDW is suppressed, three different superconducting states have been reported~\cite{Guo2022}.

One could assume that the absence of Te...Te inter-layer interactions would make single-layers of group V transition metal ditellurides more similar to disulfide and diselenides. However, this is not the case: intra-layer Te...Te interactions still make tellurides very different. Recent works on single- or few-layer NbTe$_2$ and TaTe$_2$~\cite{Bai2023,Taguchi2023,Hwang2022,DiBernardo2023,Stonemeyer2022,Feng2024} have made clear that they exhibit different and more complex structures than those of the corresponding sulfides and selenides. In addition, both the NbTe$_2$ and TaTe$_2$ single-layers have been found to occur with several different structures, some of which having modulations with large unit cells as for instance (4 $\times$ 4)~\cite{Zhang2022}, ($\sqrt19 \times \sqrt19$)~\cite{Hwang2022} and even ($\sqrt28 \times \sqrt28$)~\cite{Bai2023}. Apparently, the energy differences between the different structures are not large enough and different phases or even phase coexistence can be prepared. This is a remarkable difference with sulfides and selenides.

There have been several theoretical studies concerning TaTe$_2$ bulk~\cite{Jiang2021,Sharma2002,Mitsuichi2024,Siddiqui2021,Gao2018,Petkov2020,Liu2016} and single layers~\cite{Miller2018,Zhang2017,DiBernardo2023,Hwang2022,Feng2024}, although many of them deal either with the well-known 1$T$, 1$T$' and (3$\times$3) structures, the star-of-David structure of TaSe$_2$ or the particular structures suggested by the associated experimental studies. Comparative studies of a broad range of possible structures trying to unravel the structural and electronic factors favoring each structure are notoriously lacking. This is a serious handicap since recent works have clearly shown that single-layers or few-layers thick TaTe$_2$ phases that do not correspond to the more stable one can be indeed prepared and characterized~\cite{DiBernardo2023,Hwang2022,Feng2024}.

Although the non-occurrence of the (3$\times$3) structure for NbTe$_2$ and the absence of superconductivity at ambient pressure in TaTe$_2$ suggests some fundamental difference between NbTe$_2$ and TaTe$_2$, the previous observation points also the need to consider if different structures, characterized or proposed for the other group V MTe$_2$ single-layers, are likely to be observed for TaTe$_2$. Whereas some of the considered structures are normal metals, other display interesting electronic structures around the Fermi level and may exhibit unconventional properties. For instance, a recent theoretical study on single-layer NbTe$_2$ suggested that the ground state could be a (4$\times$4) CDW non trivial topological insulator~\cite{Zhang2022} and later a similar result was proposed for hole-doped TaTe$_2$ \cite{Wang2024}.  From the viewpoint of the chemical bonding, TaTe$_2$, like the other group V ditellurides, does not behave as expected for an octahedrally coordinated $d^1$ transition metal atom in hexagonal-based layers. In fact, understanding the occurrence of the trimerized double zigzag chains (Fig.~\ref{fig:struct_1}b) has been a challenge for longtime~\cite{Whangbo1992,Chen2018}. 

Under such circumstances, a general theoretical study of TaTe$_2$ single-layers where all structures are considered on the same footing and where both structural, electronic and bonding features are compared would be of major value. In particular we would like to understand the following questions: what distinguishes 1$T$-TaTe$_2$ from the related disulfide and diselenide?, what are the more stable structural modulations for TaTe$_2$ single-layers?, is Fermi surface nesting at the origin of some of these CDW instabilities?, how the CDW affects the metallic properties of the non distorted lattice?, can some of these CDW phases exhibit exotic properties? In this work, we report a study of a large series of possible CDW structures for 1$T$-type TaTe$_2$ single-layers and we develop a detailed analysis of the so far unclear relationship between structure and bonding. A very rich structural and electronic landscape emerges which we hope can provide useful guidelines to stabilize different TaTe$_2$ CDW.  

\begin{table}[!b]
    \centering
    \caption{Relative energies (meV/f.u.) of the 1$T$ , (3$\times$1) (i.e. 1$T$') and (3$\times$3) bulk phases of MTe$_2$ (M= V, Nb and Ta).}
    \begin{tabular}{ccccccccccc} 
    \hline
    \hline   
     &&& (3$\times$3) &&& (3$\times$1)&& 1$T$ \\
    \hline 
    TaTe$_2$  &&&  0  &&&  +6 &&  +161 \\
    NbTe$_2$  &&&  +1.8  &&&  0 &&  +134 \\
    VTe$_2$   &&&  +1.4  &&&  0 &&  +95 \\                             
    \hline 
    \hline
    \end{tabular}
    \label{tab:energies_bulk}
\end{table}

\section{Results and discussion}
\subsection{Bulk}\label{sec:results}

Before proceeding to the study of single-layers it is imperative to test the performance of the computational settings by first looking at the known structural properties of bulk MTe$_2$ (M= Ta, Nb and V). The ground state for TaTe$_2$ is found to be the (3$\times$3) phase (see Table~\ref{tab:energies_bulk}), with the (3$\times$1) one being 6 meV/f.u. (f.u.: formula unit) higher in energy. The stability is reversed for bulk NbTe$_2$ and VTe$_2$ where the (3$\times$1) phase is calculated to be more stable by 1.8 and 1.4 meV/f.u, respectively.  The ideal hexagonal 1$T$ phase of bulk TaTe$_2$ is found at considerably higher energy (160 meV/f.u.). These results are in good qualitative agreement with the experimental observation that bulk TaTe$_2$ crystals are found in the distorted (3$\times$1) monoclinic phase at room temperature (RT) but distorts towards a (3$\times$3) modulated structure at 170 K \cite{Sorgel2006}, whereas NbTe$_2$ and VTe$_2$ exhibit the (3$\times$1) monoclinic structure at all temperatures as well as with the non observation, up to date, of the 1$T$ structure for NbTe$_2$ and TaTe$_2$. However, note that for M= V the instability of the 1$T$ phase has notably decreased, which is also consistent with the fact that 1$T$ VTe$_2$ has been actually observed at high temperature $\sim$474 K \cite{Bronsema1984}. We thus conclude that the relative weight of the M-M, Te...Te and M-Te bonding is reasonably well taken into account in the present calculations thus lending credit to our calculations for the TaTe$_2$ single-layers.

\begin{figure*}[!t]
    \centering
    \includegraphics[scale=1]{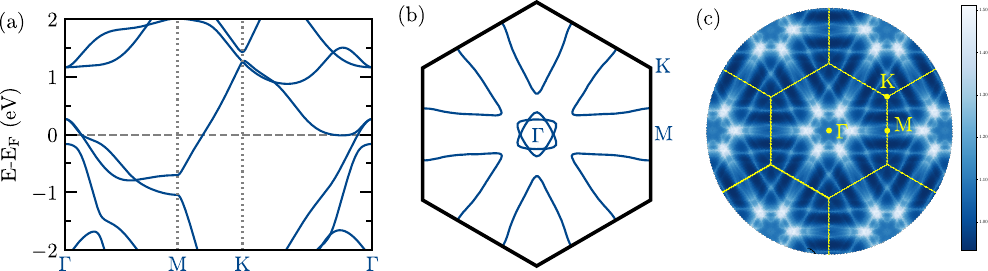}
    \caption{Band structure (a), Fermi surface (b) and Lindhard response function (c) for single-layer 1$T$-TaTe$_2$ where $\Gamma$= (0, 0), M= (1/2, 0) and K= (1/3, 1/3) in units of the reciprocal lattice vectors (see Fig.~\ref{fig:struct_1}d).}
    \label{fig:fig2}
\end{figure*}

\subsection{Single-layers}

In the following, we discuss a series of fourteen different structures (Table~\ref{tab:energies})  for single-layer TaTe$_2$ which we believe provides useful playground to understand the structural, electronic and bonding features of this system which is the subject of a large contemporary interest. Note that from now on, except when otherwise stated, we will refer to the 1$T$' structure as 3$\times$1.

\subsubsection{1\texorpdfstring{$T$}~, (3\texorpdfstring{$\times$1)}~ and (3\texorpdfstring{$\times$3)}~ structures}\label{sec:3x1_3x3}

\begin{table}[!ht]
    \centering
    \caption{Relative energies (meV/f.u.) for the different single-layer TaTe$_2$ structures considered in this work.}
 \begin{tabular}{lcccc} 
    \hline
    \hline 
     & &&&$\Delta\mathrm{E}$ (meV/f.u.) \\
     \hline
     1T               &&&& +131 \\
    \hline 
    ($3\times3$)     &&&&   0 \\
    ($3\times1$)                              &&&&  +12     \\
    \hline
    ($\sqrt{7}\times\sqrt{7}$)               &&&& +82   \\
    ($\sqrt{13}\times\sqrt{13}$)              &&&& +17   \\
    ($\sqrt{19}\times\sqrt{19}$)$_\mathrm{A}$    &&&& +64   \\
     \hline
    ($\sqrt{19}\times\sqrt{19}$)$_\mathrm{B}$  &&&& +50   \\
    ($\sqrt{19}\times\sqrt{19}$)$_\mathrm{C}$   &&&& +35   \\
     \hline
    ($4\times4$)$_\mathrm{A}$                            &&&& +1    \\
    ($4\times1$)                               &&&& +20   \\
    ($4\times4$)$_\mathrm{B}$                                &&&& +25   \\
    ($4\times4$)$_\mathrm{C}$                               &&&& +114   \\
   \hline
    ($2\times \sqrt{3}$)                       &&&& +141  \\
    ($\sqrt{7}\times \sqrt{3}$)                &&&& +103  \\
    
    \hline 
    \hline
    \end{tabular}
    \label{tab:energies}
\end{table}

\begin{table}[!ht]
    \centering
    \caption{Relative energies (meV/f.u.) of the 1$T$, (3$\times$1) (i.e. 1$T$'), (3$\times$3) and 2$H$ bulk phases of TaTe$_2$ and the associated single-layers.}
    \begin{tabular}{ccccccccccc} 
    \hline
    \hline 
        &  \multicolumn{10}{c}{Bulk}      \\
       \cline{2-11}
    Phase          &&& (3$\times$3) &&& (3$\times$1)&& 1$T$  && 2$H$  \\
    
    $\Delta\mathrm{E}$ (meV/f.u.)  &&&  0  &&&  +6 && +160 && +128 \\
    \hline 
        &  \multicolumn{10}{c}{Single layer}            \\
         \cline{2-11} 
    Phase       &&&  (3$\times$3)   &&&   (3$\times$1)  &&  1$T$    &&   1$H$   \\
    $\Delta\mathrm{E}$ (meV/f.u.)     &&&  0  &&& +12 && +131 && +122   \\                                    
    \hline 
    \hline
    \end{tabular}
    \label{tab:energies_bulk_sl}
\end{table}

The results in Table~\ref{tab:energies_bulk_sl} show that the stability of the three 1$T$ related structures 1$T$, (3$\times$1) and (3$\times$3)  of the bulk is kept in single-layers. They also allow to ascertain the role of the inter-layer Te...Te interactions. For instance, the difference between the bulk octahedral ($T$) and trigonal prismatic ($H$) non-distorted phases has strongly decreased in single-layers where the two phases become comparable in stability. This is in agreement with recent reports on the simultaneous observation of the two phases as single-layers when controlling the growth conditions so as to stabilize the 1$H$ phase \cite{DiBernardo2023,Feng2024}. It is important to note that these phases have not been observed in bulk.

\begin{figure*}
    \centering
    \includegraphics[scale=1]{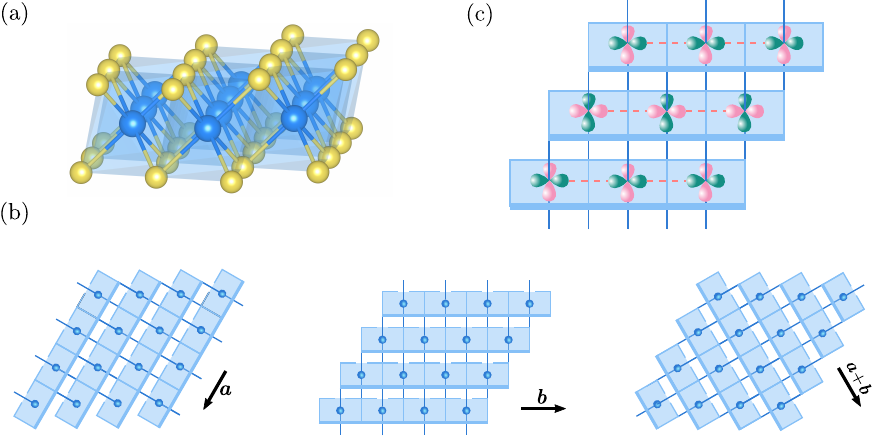}
    \caption{Octahedral layer of 1$T$-TaTe$_2$ (a) and three equivalent representations showing the occurrence of three different types of octahedral edge-sharing chains (b). (c) Schematic representation of the Ta contribution to the crystal orbital of the lowest, partially filled Ta-based band at one of the equivalent M points of the BZ.}
    \label{fig:Fig_3}
\end{figure*}

Apparently, the driving force for the 1$T$$\rightarrow$(3$\times$1) distortion is the stabilization induced by the extensive metal-metal bonding along the double zigzag chains. However, since the Ta atoms in TaTe$_2$ are in a $d^1$ configuration, the lower Ta 5$d$ band should be half-filled in the absence of Te to Ta electron transfer, so that a trimerization is difficult to explain. It is thus assumed that there must be some Te to Ta electron transfer that changes the filling of this Ta 5$d$ band \cite{Whangbo1992}. The band structure of Fig.~\ref{fig:fig2}a clearly shows the overlap between the Te 5$p$ and Ta 5$d$ bands. The two bands slightly above the Fermi level at $\Gamma$ are the Te 5$p_{x,y}$ bands and the band slightly below the Fermi level is a Ta-based band which through several avoided crossing becomes the band which crosses the Fermi level along the M$\rightarrow$K direction. Because of the overlap with the Te 5$p_{x,y}$ bands such Ta 5$d$ based band is more than half-filled. Thus, the metallic character of single-layer TaTe$_2$ is not only due to the occurrence of a half-filled Ta 5$d$ band but to the simultaneous occurrence of partially filled Ta and Te based bands. In contrast with naive expectations that suppression of the inter-layer Te...Te interactions could have a pronounced effect in the nature of the Ta-Ta clustering, the results of Table~\ref{tab:energies_bulk_sl} show that this is not really the case. 
In fact, the driving force towards the (3$\times$1) structure decreases and the tendency towards the (3$\times$3) CDW increases in the single-layer. Although the Te to Ta electron transfer certainly decreases with respect to bulk thus leading to the smaller bias for the 1$T\rightarrow$(3$\times$1) distortion, it does not strongly change. Note that, in the absence of the ``chemical pressure'' exerted by the neighboring layers in the solid, the inner structure of the octahedra in the single-layer relaxes and somewhat increases the intra-layer Te...Te contacts thus partially compensating the effect of the missing inter-layer Te...Te contacts.

In understanding the different CDW structures of 1$T$-TaTe$_2$ it is essential to understand the nature of the partially filled Ta 5$d$ band. The 1$T$-type octahedral layer (Fig.~\ref{fig:Fig_3}a) is built from the condensation of octahedra such that there are edge-shared octahedral chains along the three symmetry equivalent directions $a$, $b$ and ($a$ + $b$) (see Fig.~\ref{fig:Fig_3}b). Every Ta atom of this layer has three $t_{2g}$ orbitals, every one pointing along one of the three equivalent chain directions~\cite{Whangbo1992,Cheng2021}. Because the octahedra share edges, every Ta atom can make strong Ta-Ta $\sigma$-bonding interactions along one of these three equivalent directions. For instance, Fig.~\ref{fig:Fig_3}c shows a schematic view of the crystal orbital at one of the equivalent M points of the BZ. Because of the symmetry of the lattice there are equivalent crystal orbitals making Ta-Ta $\sigma$ bonding interactions along the three equivalent directions of the BZ. Thus, at every $k$-point, the Ta 5$d$ orbitals mix to generate a tilted Ta $d_{x^2-y^2}$-type orbital making these strong $\sigma$-type interactions along one direction and very weak $\delta$-type interactions with those of the parallel chains. To summarize, the partially filled band of TaTe$_2$ describes the formation of Ta-Ta bonds along the lattice~\cite{Whangbo1992}. As it will be seen later, the different modulated structures are different ways to stabilize the system by locally strengthening the bonding in some Ta-Ta contacts at the expense of a weakening in the other (for instance strengthening the Ta-Ta bonds of the trimerized double zigzag chains in the (3$\times$1) structure at the expense of those in-between the triple chains).

The calculated FS and the associated Lindhard response function of the 1$T$ structure are shown in Figs.~\ref{fig:fig2}b and c, respectively. In agreement with the previous discussion, the FS can be described as the superposition of three Ta based pseudo-1D portions along the $a$*, $b$* and ($a$* + $b$*) and a central part with two superposed rounded triangles which are hole pockets originating from the Te 5$p_{x,y}$-type bands around $\Gamma$. The Ta based contribution originates from the band described above (see Fig.~\ref{fig:Fig_3}). If it were not for the overlap with the Te 5$p_{x,y}$ bands there will be one electron to fill this band which would thus be half-filled. However, because of the overlap with the Te 5$p_{x,y}$ bands it is a bit more than half-filled. According to our calculations the Te to Ta transfer is around 0.1 electrons. However, this value is very easy to change: a small decrease of the $a$ cell constant raises the top of the Te 5$p_{x,y}$ band and increases the electron transfer. As it will be shown later, this feature has most likely an important role in stabilizing different superstructures of 1$T$ TaTe$_2$.

\begin{figure}[t]
    \centering
    \includegraphics[scale=1]{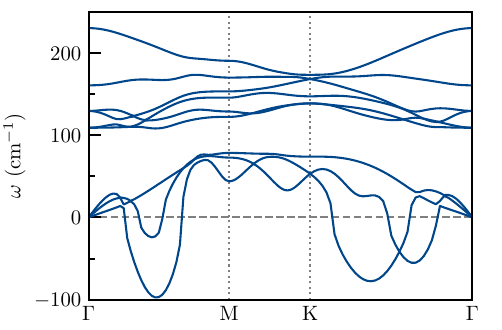}
    \caption{Phonon dispersion calculated for single-layer 1$T$-TaTe$_2$.}
    \label{fig:ep}
\end{figure}

Because of the flat portions, the FS exhibits nesting vectors which could be associated with electronic instabilities. Such possibility has been evoked in the literature and some comment is in order~\cite{Battaglia2005}. In the absence of Te to Ta electron transfer the partially filled band would be filled with one electron. As discussed above, this band can be considered to result from the superposition of three 1D chains of orbitals. Therefore, in the absence of Te to Ta electron transfer every one of these chains would be associated with one-third electron and thus its 2$k_F$ nesting vector would have a 1/6 component along the chain direction. Because of the Te to Ta electron transfer the nesting vectors should be a bit larger than this value. Indeed, as expected from the previous analysis, the Lindhard electron hole response exhibits slightly warped lines of intensity maxima perpendicular to the chains direction, i.e. $a$, $b$ and 1/2($a$+$b$). The 2$k_F$ component of these lines along the chain direction is $\sim$ 0.19, i.e. a bit larger than the 1/6 value because of the Te to Ta electron transfer. There are additional diffuse slightly warped lines because of nesting implicating  the inner part of the FS around $\Gamma$. Nevertheless, the important observation is that there are portions with maxima of the electron-hole response around the 0.19 $a$* and equivalent points of the calculated Lindhard function (Fig.~\ref{fig:fig2}c). However, they are not sharp maxima as needed for a Fermi surface driven mechanism of CDW formation; they are shallow regions around this point evoking the situation for 1$H$  and 2$H$ NbSe$_2$~\cite{Johannes2006,Guster2019}. In addition, the nesting vector is not commensurate an does not correspond to a trimerization (i.e. 1/3 component along the chain direction). Thus, a FS nesting mechanism (i.e. a weak-coupling scenario) can be dismissed as the origin of the 1$T$$\rightarrow$(3$\times$1) transition. Note that he optimized structure for the (3$\times$1) structure is 119 meV/f.u. more stable than 1$T$-TaTe$_2$. The calculated Ta-Ta bond lengths are 3.228 \AA, i.e. 0.43 \AA~shorter than in the 1$T$ phase. This is a very strong shortening (12.3 \%), even stronger than in 1$T$-TaSe$_2$ ($\sim$ 0.24 \AA) and 1$T$-TaS$_2$ ($\sim$ 0.35 \AA) \cite{Brouwer1980, Yamamoto1983, Spijkerman1997} which are classical examples of CDWs due to a strong-coupling electron-phonon based mechanism. CDWs  due to a Fermi surface nesting mechanism (i.e. weak-coupling mechanism) usually lead to bond length variations considerably smaller~\cite{Pouget2024}. We thus conclude that the driving force for the 1$T$$\rightarrow$(3$\times$1) distortion of TaTe$_2$ must be related to the electron-phonon coupling  (i.e. a strong-coupling CDW mechanism).

The calculated phonon dispersion for 1$T$-TaTe$_2$ is shown in Fig.~\ref{fig:ep}. Two imaginary phonon modes can be identified. However they do not correspond to the motion leading from 1$T$ to the (3$\times$1) superstructure. In fact the phonon dispersion of Fig.~\ref{fig:ep} suggests that the 1$T$- structure should tend to distort towards (4$\times$1) or (4$\times$4) superstructures. Nevertheless, the trimerized structures are those reported repeatedly in experimental studies of the bulk~\cite{Brown1966,Sorgel2006} or surface~\cite{Chen2018,Feng2016,Baggari2020}. As discussed below, we (see Table~\ref{tab:energies}) and others~\cite{Wang2024} have found energy minima with this type of periodicity and stability not far from that of the (3$\times$3) phase. These observations, as well as the large number of CDW modulations reported for single- and few-layer TaTe$_2$ (and NbTe$_2$), lead us to suspect that there is a peculiar mechanism operating in these 1$T$ type tellurides.

\begin{figure}
    \centering
    \includegraphics[scale=1]{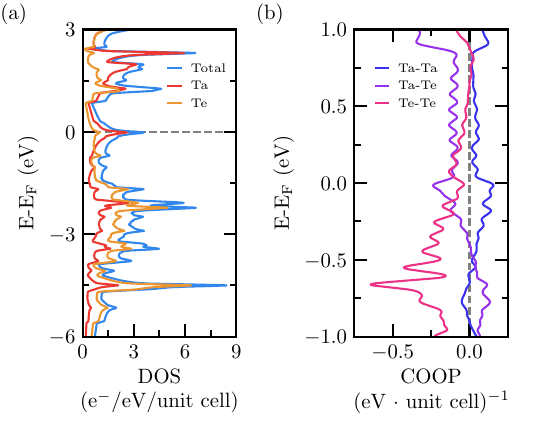}
    \caption{(a) Density of states and (b) Crystal Orbital Overlap Population (COOP) curve around the Fermi level for single-layer 1$T$-TaTe$_2$.}
    \label{fig:COOP}
\end{figure}

The best way to approach this question is by coming back to the (3$\times$1) room temperature structure which is the simplest modulation and the room temperature structure for the three group V MTe$_2$ solids (M= V, Nb, Ta). Both in bulk and single-layer the stabilization with respect to the 1$T$ ideal structure is very large (Table~\ref{tab:energies_bulk_sl}). The optimized Ta-Ta distance in the 1$T$ structure is 3.651 \AA~whereas the `a'-type bonds (see Fig.~\ref{fig:struct_1}e) in the (3$\times$1) structure are 3.228 \AA. In addition there are two different Ta-Ta contacts (`b'- and `c'-type in Fig.~\ref{fig:struct_1}e) which although larger, 3.474 \AA, have also been shortened (5 \%). A simple but useful way to discuss the bonding in solids is by considering the so-called Crystal Orbital Overlap Population plots (COOP)~\cite{Hughbanks1983,Dronskowski2005}. Taking into account all Ta-Ta pairs in the unit cell it is found that the Ta-Ta overlap population per formula unit evolves from 0.2096 to 0.3076 when going from the 1$T$ to the (3$\times$1) structures. This very large increase in metal-metal bonding (+47~\%) largely overrides the cost in Ta-Te bonding and Te...Te non-bonding interactions due to the associated distortion leading to the 119 meV/f.u. stabilization. Note that 69~\% of the Ta-Ta overlap population/f.u. is associated with the diagonal `a'-type bonds, 31~\% with the vertical `b'- and `c'-type longer Ta-Ta contacts and the contribution from the inter-chain contacts is negligible. Note that every Ta atom participates in 3 Ta-Ta interactions in 1$T$ but the average number of bonds per Ta atom in (3$\times$1) is 4/3. The large increase of Ta-Ta overlap population/f.u. is thus associated with the creation of robust Ta-Ta bonding along the diagonal directions of the chain. In short, maximization of metal-metal bonding through formation of the uncommon double zigzag chains is the driving force of the 1$T\rightarrow$(3$\times$1) distortion. However, since the Ta atoms are in a $d^1$ configuration, the Ta-Ta bonds of this double zigzag chain can not be usual two-electron two-center bonds since there are not enough electrons (three electrons per repeat unit of the double zigzag chain). This is very clear when noting that the standard Ta-Ta single bond is $\sim$2.92~\AA~\cite{Pyykko2009}, which is shorter than the Ta-Ta bond lengths in this phase.

\begin{figure}[t]
    \centering
    \includegraphics[scale=1]{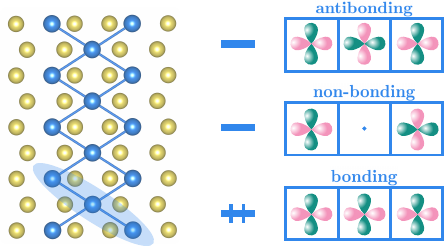}
    \caption{Three-center two-electron $\sigma$ bonding scheme for (3$\times$1)-TaTe$_2$ (right) and schematic representation of the double zigzag chain (left), with one of the two three-centre two-electron $\sigma$-bonds per repeat unit highlighted.}
    \label{fig:3c_2e}
\end{figure}

The density of states of 1$T$ TaTe$_2$ around the Fermi level (Fig.~\ref{fig:COOP}a) is quite heavily based on Ta orbitals and results from the overlap of a Ta-based band and the top of the Te 5$p_{x,y}$ bands (see Fig.~\ref{fig:fig2}a). As shown in the COOP curve of Fig.~\ref{fig:COOP}b, even with the long Ta-Ta distance of 1$T$-TaTe$_2$ (3.651 \AA), the states of the Ta band just above the Fermi level are Ta-Ta bonding whereas those of the Te bands just below the Fermi level are Te...Te antibonding. Therefore, destabilizing the Te bands (for instance by slight contraction of the cell constants) will induce an electron transfer from Te toward the Ta band. Consequently, the transferred electrons will populate states which are Ta-Ta bonding and will induce the shortening of some Ta-Ta contacts of the hexagonal 1$T$ lattice. The question is why the system chooses to strengthen the Ta-Ta contacts leading to the  peculiar trimerized double zigzag chains?

\begin{figure*}[t]
    \centering
    \includegraphics[scale=0.95]{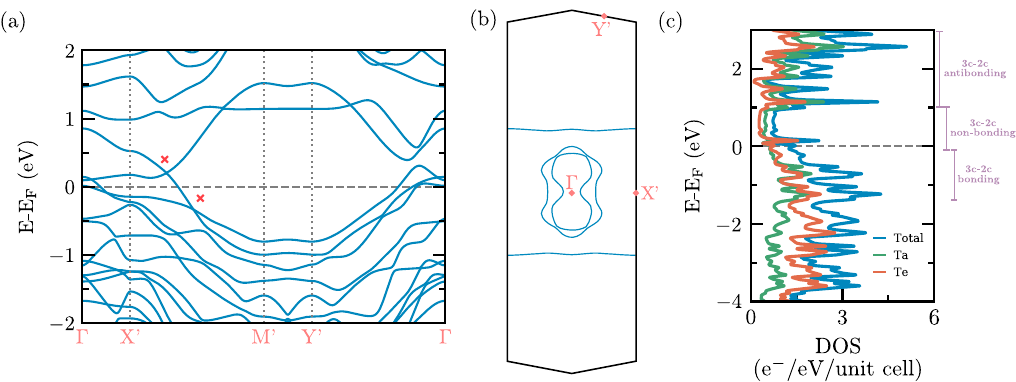}
    \caption{Band structure (a), Fermi surface (b) and Density of States (DOS) (c) calculated for (3$\times$1)-TaTe$_2$ (see Fig.~\ref{fig:struct_1}d). The energy region where the three 3c-2e levels occur are noted in (c). $\Gamma$, X', Y' and M' refer to the (0, 0), ($a$'*/2, 0), (0, $b$'*/2) and ($a$'*, $b$'*/2) points of the BZ where $a$'= $a_c$= 1/2($a$-$b$) and $b$'= $b$ (see Figs.~\ref{fig:struct_1}b and d).}
    \label{fig:3x1_BS_FS}
\end{figure*}

The answer is simple if the Te to Ta electron transfer is not far from 1/3 electron per Ta atom. In that case, the Ta atoms will be in a formal $d^{4/3}$ configuration and, since the repeat unit of the double zigzag chain contains three Ta atoms, there will be four electrons to fill Ta-Ta bonding states. As mentioned above (see also Fig.~\ref{fig:Fig_3}), the octahedra of a 1$T$ layer share their edges and thus, local chemical bonds can be established using the $t_{2g}$ orbital set, i.e. the in-plane orbital pointing toward these edges (see Fig.~\ref{fig:Fig_3}). Using one of the three Ta $t_{2g}$ orbitals for a set of three consecutive octahedra linearly arranged, the set of three orbitals shown in Fig.~\ref{fig:3c_2e} can be built. The middle orbital is metal-metal non-bonding and hence it is not essential that this level is filled to insure the stability of the system. When two electrons fill the lowest of this set of three levels, a so-called three-centre two electron (3c-2e) $\sigma$-bond is created. This type of bond is frequently found and well characterized in the chemistry of electron deficient systems like borides or transition metal clusters~\cite{Albright2013}. Of course such bonds are not as strong as the usual two-center two-electron bonds. As mentioned, the usual Ta-Ta $\sigma$ bond length is $\sim$2.92 \AA~while the bond length in the double zigzag chain is $\sim$3.23 \AA. As a matter of fact, the repeat unit of the double zigzag chain contains two sets of three longitudinally arranged octahedra, i.e. the double zigzag chain can be considered as a succession of two sets of tilted 3c-2e bonds along the chain. Thus, if the Te to Ta electron transfer is around 1/3 per Ta the electronic requirement to stabilize the (3$\times$1) structure (i.e. four electrons to fill the bonding level of the two sets of tilted 3c-2e bonds of the repeat unit) is fulfilled. Of course, in the solid the three orbitals of Fig.~\ref{fig:3c_2e} broaden into bands which can be located through the combined use of DOS and COHP or COOP curves~\cite{Dronskowski2005}. Shown in Fig.~\ref{fig:3x1_BS_FS}c is the DOS of (3$\times$1)-TaTe$_2$ where the energy region where the three types of 3c-2e levels occur have been highlighted. The bonding levels are largely filled in agreement with the 3c-2e model (see Fig. S1 in the Supplemental Material (SM)). Note that the formation of trimeric bonding states, originally proposed by Whangbo and Canadell~\cite{Whangbo1992}, was substantiated through low energy electron diffraction intensity versus voltage (LEED $I$-$V$)~\cite{Chen2018} and high resolution angle-resolved photoemission spectroscopy (ARPES) measurements~\cite{Mitsuichi2024} on surface 1$T$-TaTe$_2$.

The calculated band structure and Fermi surface for (3$\times$1)-TaTe$_2$ are shown in Fig.~\ref{fig:3x1_BS_FS}. The (3$\times$1) phase is found to be metallic in agreement with the conductivity measurements. The FS of Fig.~\ref{fig:3x1_BS_FS}b can be easily related to that of the original 1$T$ structure when the relationship of the two BZ shown in Fig.~\ref{fig:struct_1}d is taken into account. It contains closed portions around $\Gamma$ mostly originating from Te 5$p$ orbitals although hybridized with Ta 5$d$ orbitals. The more interesting feature of this FS is the occurrence of a purely one-dimensional (1D) contribution. According to our calculations, the pair of very flat contributions are almost perfectly nested by a 0.34 $b$* wave vector. In addition, note that two of the 1D components of the FS of the 1$T$ structure (see the two 'diagonal' components in Fig.~\ref{fig:struct_1}b) have disappeared. This means that the above mentioned Ta $t_{2g}$ orbitals along the diagonal directions have been completely removed from the FS by the distortion and they have been used to make the 2c-3e bonds discussed above. In addition, the severe distortion associated with the Ta-Ta bonding along the zigzag chains has induced a noticeable redistribution of the electrons at the FS such that the only remaining 1D component is now associated with an almost exactly 1/3 empty  band. Note that such 1D component of the FS has been experimentally characterized for bulk (3$\times$1)-TaTe$_2$ by ARPES although for a slightly different nesting vector, presumably because of the different Te to Ta electron transfer in bulk and single-layer~\cite{Mitsuichi2024,Lin2022}. The observation of this open FS component suggests that although single-layer TaTe$_2$ should be a two-dimensional (2D) metal it has a strong 1D metallic component which makes it susceptible to a trimerizetion along the $b$ direction. Such nesting-based trimerization would mostly destroy the 1D component of the FS but the closed portions around $\Gamma$ would remain. Consequently, our calculations anticipate the occurrence of a (3$\times$1)$\rightarrow$(3$\times$3) structural modulation that would keep the metallic character of the system. This observation, which seems to be consistent with the occurrence of a transition towards a (3$\times$3) metallic phase around 170 K in bulk (3$\times$1)-TaTe$_2$~\cite{Sorgel2006}, deserves, however, further discussion.

\begin{figure*}[t]
    \centering
    \includegraphics[scale=1]{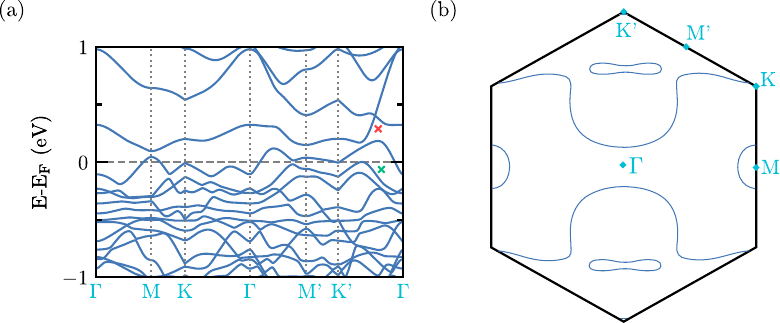}
    \caption{Band structure (a) and Fermi surface (b) calculated for single-layer (3$\times$3)-TaTe$_2$.
    $\Gamma$, M, K, M' and K' refer respectively to the (0, 0), ($a$'*/2, 0), ($a$'*/3, $b$'*/3), (0, $b$'*/2) and (-$a$'*/3, 2$b$'*/3) points of the BZ where $a$'= $a'_{c}$= 1/2($a$-$b$) and $b$'= $b$ (see Figs.~\ref{fig:struct_1}c and d).}
    \label{fig:3x3_BS_FS}
\end{figure*}

The origin of the (3$\times$1)$\rightarrow$(3$\times$3) transition in bulk TaTe$_2$ has been considered in several recent publications~\cite{Mitsuichi2024,Siddiqui2021,Gao2018,Petkov2020,Lin2022}. In the context of the present work, the two above mentioned ARPES studies are specially relevant~\cite{Lin2022,Mitsuichi2024}. In both of these works, 1D states were observed for the (3$\times$1) phase. However, the wave vector of the associated nesting vector does not coincide with that needed to justify the threefold increase of the unit cell. From this it was concluded that a weak-coupling nesting scenario cannot be at the origin of the transition which should be related to some phonon based instability. We note that our calculations for bulk (3$\times$1)-TaTe$_2$ agree with these results in that the 1D component is quite heavily warped and does not justify the threefold increase of the periodicity along the second in-plane direction (see Fig. S2 in the SM). Nevertheless, as we have said before, the same calculations for the single-layer (see Fig.~\ref{fig:3x3_BS_FS}b)  led to a very flat 1D component very near the 1/3$b$* wave vector. Altogether, these observations suggest that there is an important effect of the reduction of screening effects in the single layer. However, since the evolution towards the (3$\times$3) structure occurs both for bulk and single-layer and the nesting vector differs in the two situations (and additionally is noticeably warped for the bulk but non warped for the single-layer), it would be very surprising that FS nesting was really the driving force for the transition. For the reasons noted below a strong-coupling phonon based mechanism is a more rational alternative. The calculated band structure and FS of single-layer (3$\times$3) TaTe$_2$ is shown in Fig.~\ref{fig:3x3_BS_FS} (see Fig.~\ref{fig:struct_1}d for he relationship between the (3$\times$1) and (3$\times$3) BZ). Note that the FS of Fig.~\ref{fig:3x3_BS_FS}b still contains remnants of the (3$\times$1) nested FS which originate from a strongly avoided crossing between the bands noted with green and red crosses in Fig.~\ref{fig:3x3_BS_FS}a. This component just touches the Fermi level and can occur or vanish as a function of very minor structural changes that practically do not affect the calculated total energy. Thus, nesting can not be the main source of stabilization of the (3$\times$3) structure and the structural analysis of next paragraph will reinforce this conclusion.

When distorting from 1$T$ to (3$\times$1) the Ta atoms have used only two of the three $t_{2g}$ orbitals to make the 3c-2e bonds. As a consequence, two of the 1D components of the FS have been removed. One of the three $t_{2g}$ orbitals, that pointing along the $b$ direction, essentially has not been used. These orbitals have a strong contribution to the dispersive band of single-layer (3$\times$1)-TaTe$_2$ (noted with a red cross in Fig.~\ref{fig:3x1_BS_FS}a) which after undergoing a slightly avoided crossing near the Fermi level becomes almost 2/3 filled. The driving force for the distortion is the attempt to use these orbitals to increase the Ta-Ta bonding. Our optimized structure clearly reflects this feature: the uniform double zigzag chain evolves towards a kind of trimerized chain with `clusters' of seven Ta atoms with two strongly shortened inner Ta-Ta bonds ($\Delta$= -0.24 \AA, `b' bonds in Fig.~\ref{fig:struct_1}e) and a broken one ($\Delta$= +0.75 \AA, `e' bond in Fig.~\ref{fig:struct_1}e). The formation of the two additional bonds essentially needs a displacement of the two outer Ta atoms in opposite directions toward a central one. In that way the two `a' bonds become shorter ($\Delta$= -0.15 \AA), the two `f' type ones become longer ($\Delta$= +0.19 \AA) and the `d' ones remain almost unchanged. Using the $t_{2g}$ orbital which does not participate in the `diagonal' 3c-2e bonding to increase the Ta-Ta bonding thus needs a substantial electronic reorganization around the two outer Ta atoms of the `cluster' so that the total energy gain can not be large (see Table~\ref{tab:energies_bulk_sl}). As a matter of fact the total Ta-Ta overlap population/f.u. remains practically unaltered (the variation is smaller than 1\% with respect to the (3$\times$1) phase). However, the reorganization of the Ta-Ta network brings about an increase of the Ta-Te and Te...Te overlap populations which definitely stabilizes the structure. In short, using the third $t_{2g}$ orbital of the central Ta atoms induces a complex reorganization keeping almost unaltered the stability of the Ta-Ta network but creating a less strained situation stabilizing the Ta-Te and Te...Te networks. Since the total Ta-Ta bonding is essentially kept, the energy gain of the (3$\times$1)$\rightarrow$(3$\times$3) process is only modest, around one order of magnitude weaker than the stabilization afforded in the formation of the (3$\times$1) structure. Nonetheless, the atomic displacements are again found to be large and the final structure can hardly be seen as a slight structural modulation of the initial one as it would be for a weak-coupling nesting based scenario. Even if the energy gain is relatively small, the associated structural rearrangement is very substantial and this translates into a strong orbital hybridization of the bands around the Fermi level which are transition metal based. 

Let us note that it has been frequently considered as puzzling that the increase of conductivity below the transition is surprising for a CDW material. This comment would be appropriate if FS nesting was the driving force for the transition towards the (3$\times$3) structure because in that process the FS is partially or totally removed. Nevertheless, in a strong-coupling scenario the FS is only modified because of the structural changes induced by the phonon instability and the conductivity may either decrease or increase. NbTe$_4$ and TaTe$_4$ provide related examples of strong-coupling CDW materials where there is practically no change in conductivity through the CDW transition~\cite{Guster2022}.  

\begin{figure*}
    \centering   \includegraphics[scale=0.9]{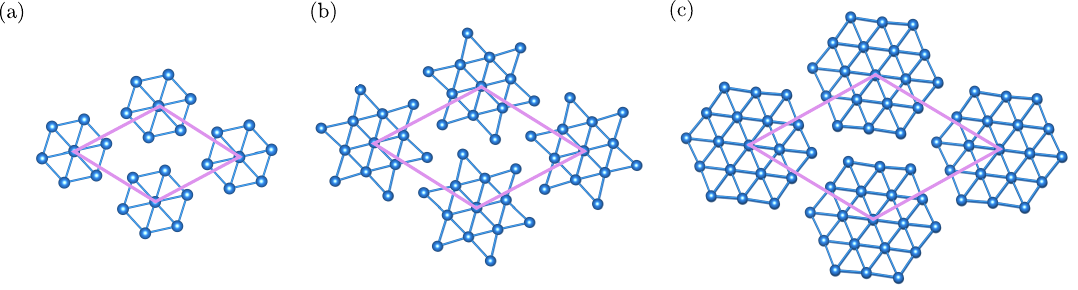}
    \caption{Different ($\sqrt m\times \sqrt m$) single-layer TaTe$_2$ structures optimized in this work: ($\sqrt 7\times \sqrt 7$) (a), ($\sqrt{13}\times \sqrt{ 13}$) (b), and ($\sqrt 19\times \sqrt 19$) (c). Only the Ta atoms are shown for simplicity. The last phase is refereed to as ($\sqrt{19}\times \sqrt{19}$)$_\mathrm{A}$ in the text to distinguish it from other ($\sqrt {19}\times \sqrt {19}$) phases.}
    \label{fig:struct_2}
\end{figure*}

In short, the formation of the (3$\times$3) structure from the ideal hexagonal 1$T$ one can be seen as a two-step process in which a set of three trimerizations along the main directions of the hexagonal lattice occur, each of them using one of the three Ta $t_{2g}$ orbitals. However, using just two of the $t_{2g}$ orbitals in the 1$T\rightarrow$(3$\times$1) transition is very efficient in lowering the energy of the system because chains with a very substantial metal-metal bonding are created. The third trimerization using the remaining $t_{2g}$ orbitals, allows to reach a structure with a more comfortable situation for the Ta-Te and Te...Te networks without losing the stabilization of the metal-metal network. Although structurally significant, the second transition brings about only a weak stabilization which should be easily tuned by physical and chemical means. In fact, it is not surprising that the such (3$\times$1)$\rightarrow$(3$\times$3) transition only occurs for TaTe$_2$ since Ta has more extended $d$ orbitals making the reorganization of the transition metal network easier.

\subsubsection{\texorpdfstring{($\sqrt m\times \sqrt m$)} ~ structures}\label{sec:stars}

In the previous section we analyzed the results in considerable detail because in that way we introduced the main ideas needed for understanding the nature and stability of different TaTe$_2$ CDWs. Essentially the same ideas can be used to understand the results of this and the next sections so that the discussion will be more succinct and only the more relevant aspects will be commented. In this section we report the results concerning TaTe$_2$ single-layers containing exclusively $\sqrt m\times \sqrt m$ Ta clusters with $m$= 7, 13 and 19.

The ($\sqrt 13\times \sqrt 13$) CDW (Fig.~\ref{fig:struct_2}b) is the ground state for bulk 1$T$-TaSe$_2$ and 1$T$-TaS$_2$ bulk \cite{Wilson1975, Scruby1975} and single-layer \cite{Chen2020,Lin2020} but it has neither been observed for bulk nor single-layer TaTe$_2$ although it has been proposed that ($\sqrt 13\times \sqrt 13$) and (3$\times$3) CDW may coexist in few-layers TaTe$_2$~\cite{Hwang2022}. 
The ($\sqrt {19}\times \sqrt {19}$) CDW phase of Fig.~\ref{fig:struct_2}c was proposed longtime ago for bulk TaTe$_2$ and NbTe$_2$ \cite{Landuyt1974,Landuyt1978} although the detailed structure was never reported.
The inner 7-atom core of the ($\sqrt 13\times \sqrt 13$) cluster has been proposed as being part of some few-layer CDW phases as the metastable ($\sqrt {19}\times \sqrt {19}$) CDW phase recently reported~\cite{Hwang2022}.
Furthermore, single- or few--layer ($\sqrt 19\times \sqrt 19$) CDW (and even higher members of the family as ($\sqrt 28\times \sqrt 28$) \cite{Bai2023}) have been recently reported for TaTe$_2$ and NbTe$_2$ \cite{Bai2023,Hwang2022,DiBernardo2023}. 
Nevertheless, it is not clear (see below) that they contain the nineteen atom clusters of Fig.~\ref{fig:struct_2}c.
Interestingly, the pure ($\sqrt 7\times \sqrt 7$) CDW (Fig.~\ref{fig:struct_2}a) phase has not been observed as single-layer.

As shown in Table~\ref{tab:energies}, the ($\sqrt 13\times \sqrt 13$) phase is only slightly higher in energy than the very stable (3$\times$1) phase, in agreement with the suggestion that it could coexist with the (3$\times$3) phase in few-layers thick films \cite{Hwang2022}.  The ($\sqrt 7\times \sqrt 7$) and ($\sqrt {19}\times \sqrt {19}$)$_\mathrm{A}$ CDW are noticeably less stable. However, this does not mean that they can not be stabilized in single- or few-layers TaTe$_2$. Note that the ($\sqrt {19}\times \sqrt {19}$) phase prepared by Hwang $et~al.$ \cite{Hwang2022}, which will be refereed to as ($\sqrt {19}\times \sqrt {19}$)$_\mathrm{B}$, is already noticeably less stable than the (3$\times$3) and (3$\times$1) ones (see Table~\ref{tab:energies}) and yet it could be grown and partly characterized under suitable conditions. 

\begin{figure}[t]
   \centering
    \includegraphics[scale=1]{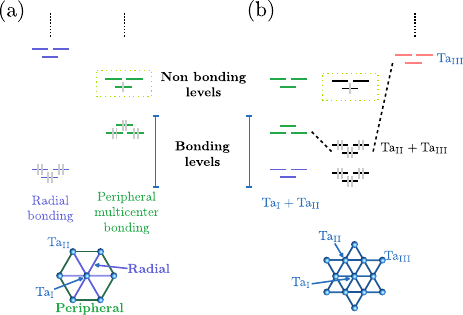}
    \caption{Schematic description of multicenter bonding for a 13-atom star-of-David cluster \cite{Whangbo1992}.}
    \label{fig:star_scheme}
\end{figure}

The very stable ($\sqrt {13}\times \sqrt {13}$) phase contains three different Ta atoms: (i) that at the center of the star-of-David cluster (Ta$_{\mathrm{I}}$), (ii) the six Ta atoms in the hexagon around Ta$_\mathrm{I}$ (Ta$_{\mathrm{II}}$), and the six Ta atoms capping this hexagon (Ta$_{\mathrm{III}}$) (see Fig.~\ref{fig:star_scheme}b). All Ta$_\mathrm{I}$-Ta$_{\mathrm{II}}$ and Ta$_{\mathrm{II}}$-Ta$_{\mathrm{II}}$ distances, i.e. all Ta-Ta bonds within the 7-atoms core of the cluster, are 3.26 \AA, which are only 1\% longer than in the very stable (3$\times$1) phase. The Ta$_\mathrm{II}$-Ta$_\mathrm{III}$ bonds are in average 3.505 which are noticeably longer but yet shorter than in the 1$T$ phase (3.651 \AA). Thus, the central 7-atoms cluster must provide a strong stability to this phase. The total Ta-Ta overlap population/f.u., i.e. the average Ta-Ta overlap population, is calculated to be only 15.2\% smaller than in the very stable (3$\times$1) phase. It is then normal that the ($\sqrt {13}\times \sqrt {13}$) phase is only 5 meV/f.u. less stable than the (3$\times$1) phase because of the extensive metal-metal bonding. Further analysis of the Ta-Ta bonding shows that 57\% of the Ta-Ta overlap population/f.u. is due to the Ta$_\mathrm{I}$-Ta$_\mathrm{II}$ and Ta$_\mathrm{II}$-Ta$_\mathrm{II}$ bonds of the 7-atom core, 33\% is due to the Ta$_\mathrm{II}$-Ta$_\mathrm{III}$ outer bonds of the star-of-David cluster, and finally 10\% to the Ta-Ta contacts between clusters.

As for the (3$\times$1) phase, the extensive Ta-Ta bonding can not be due to usual 2 center-2 electron bonds because there are only 13 electrons/f.u. to fill the Ta-based bands and some kind of electron deficient bonding scheme must be at work. The essential ideas to understand the bonding in such kind of clusters were developed by Whangbo and Canadell\cite{Whangbo1992} and can be easily adapted to the present system (Fig.~\ref{fig:star_scheme}). Essentially, the model relies on the following ideas. (1) The Ta$_\mathrm{I}$-Ta$_\mathrm{II}$ and Ta$_\mathrm{II}$-Ta$_\mathrm{II}$ bonds of the 7-atom core are considerably shorter than the Ta$_\mathrm{II}$-Ta$_\mathrm{III}$ ones. Consequently, we can consider the 13-atom cluster as arising from a central 7-atom cluster (Ta$_\mathrm{I}$ and Ta$_\mathrm{II}$) and a group of six outer (i.e. capping) Ta$_\mathrm{III}$ atoms which simply ‘follow’ the distortion of the 7-atom cluster core. Within this core there are two different types of Ta-Ta bonds: (i) radial (Ta$_\mathrm{I}$-Ta$_\mathrm{II}$) and (ii) peripheral (Ta$_\mathrm{II}$-Ta$_\mathrm{II}$) (see Fig.~\ref{fig:star_scheme}). (2) The central Ta$_\mathrm{I}$ atom participates in three separate linear 3c-2e $\sigma$-bonding interactions with the `peripheral' atoms, i.e. the ‘radial’ bonds in Fig.~\ref{fig:star_scheme} thus involve six electrons. (3) As a consequence, the distance between the six peripheral atoms is shortened and they can use a ‘tangential’ $d$ orbital to make bonding interactions within the outer six-membered ring, i.e. the ‘peripheral’ interactions in Fig.~\ref{fig:star_scheme}. From the six peripheral $d$ orbitals, six symmetry-adapted combinations can be built, three of which are bonding and three antibonding, describing the multicenter bonding in the periphery of the 7-atom cluster. Thus, filling the three bonding states will insure the peripheral bonding of the cluster. Consequently,  the low-lying three bonding 3c-2e radial orbitals and the three multi-centre bonding peripheral orbitals insure the bonding requirements of the central 7-atom cluster which thus requires 12 bonding electrons. (4) The $t_{2g}$ levels of the six Ta$_\mathrm{III}$ atoms are higher in energy in than those of the Ta$_\mathrm{I}$ and Ta$_\mathrm{II}$ because of the stronger local octahedral distortion \cite{Canadell1991} and thus act simply as providing an additional stability to the twelve filled orbitals describing the bonding of the 7-atoms core, and somewhat delocalizing these levels to the outer part of the star-of-David cluster. (5) Consequently, 12 of the 13 electrons of the star-of-David cluster are used to fill the set of six bonding orbitals (three radial and three peripheral) and the remaining electron fills the lower antibonding tangential orbital(s) which are essentially Ta-Ta non bonding so that the formation of the 13-atoms cluster is energetically favourable. 

\begin{figure*}[t]
    \centering
    \includegraphics[scale=1]{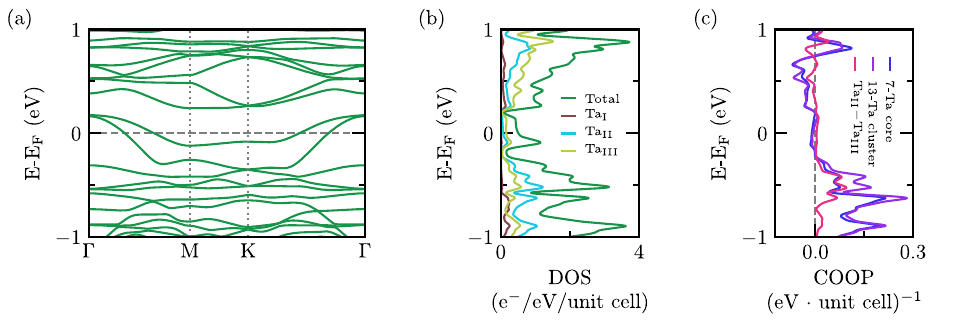}
   \caption{Electronic structure of the single-layer ($\sqrt {13}\times \sqrt {13}$) phase. (a) Band structure where $\Gamma$= (0, 0), M= (1/2, 0) and K= (1/3, 1/3) in units of the reciprocal lattice vectors. (b) Total Density of States and partial contributions of the different Ta atoms. (b) COOP curves for the total Ta-Ta bonding within the 13-atom clusters, the Ta-Ta bonding within the inner 7-atoms core (Ta$_\mathrm{I}$-Ta$_\mathrm{II}$ and Ta$_\mathrm{II}$-Ta$_\mathrm{II}$ bonds), and the Ta-Ta bonding in the outer portion of the 13-atom cluster (Ta$_\mathrm{II}$-Ta$_\mathrm{III}$ bonds).}
    \label{fig:star_13}
\end{figure*}

\begin{figure*}
    \centering
    \includegraphics[scale=1]{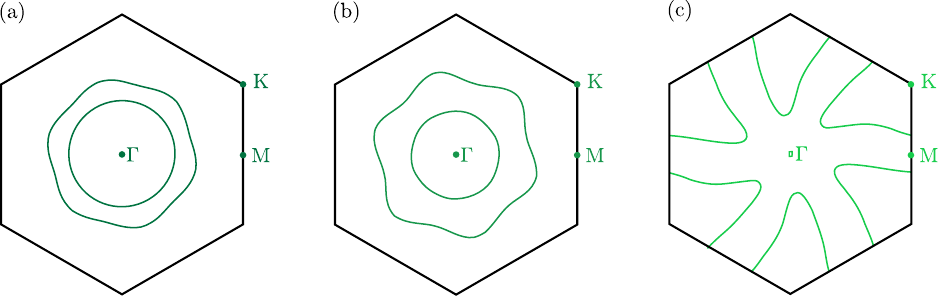}
    \caption{Fermi surfaces calculated for the single-layer phases ($\sqrt 7\times \sqrt 7$) (a), ($\sqrt {13}\times \sqrt {13}$) (b) and ($\sqrt {19}\times \sqrt {19}$)$_\mathrm{A}$ (c).}
    \label{fig:stars_FS}
\end{figure*}

This scheme is fully supported by the present DFT calculations (Fig.~\ref{fig:star_13}). The Fermi level cuts two bands which globally contain one electron. The associate FS is typical of a 2D metal (see Fig.~\ref{fig:stars_FS}). These bands even if containing a large contribution of the Ta atoms are Ta-Ta non bonding (see Figs.~\ref{fig:star_13}b and c) as predicted by the qualitative model. Below, in the region between -0.25 and -0.70 eV, the levels largely associated with the peripheral bonding are found. These levels are those which preferentially mix in character from the outer Ta$_\mathrm{III}$ atoms. Below, in the energy range between -0.6 and -1.1 eV, the levels largely associated with the radial bonding are found. Of course, some interaction and mixing occurs between the two sets of three bands but the main character with the radial bonding levels below the peripheral ones is kept (see Fig. S3 in SM). The radial and peripheral contributions to the Ta-Ta bonding of the inner 7-atom core of the cluster are similar, i.e., 53.4 \% and 46.6\%, respectively.  
The more noticeable difference with the sulfide and selenide counterparts is that the non bonding level is now a set of two bands with both Ta$_\mathrm{II}$ and Ta$_\mathrm{III}$ character whereas in the sulfide and selenide cases is a single band with mostly Ta$_\mathrm{I}$ and Ta$_\mathrm{II}$ character \cite{Whangbo1992,Cheng2021}. Although in both cases these levels are Ta-Ta non bonding in character, there is an important difference: in the sulfides and selenides the band is very flat because is mostly concentrated in the central Ta$_\mathrm{I}$ and Ta$_\mathrm{II}$ 5$d$ orbitals. However, in the present case the partially filled bands are substantially dispersive because they are not mostly confined into the central Ta$_\mathrm{I}$ and Ta$_\mathrm{II}$ 5$d$ orbitals. Thus, even if the TaTe$_2$ ($\sqrt {13}\times \sqrt {13}$) phase can be stabilized it is not expected that the peculiar Mott type behavior of the sulfide and selenide phases will occur. Another important observation is that with the electron counting of this phase the Ta-Ta bonding is kept practically constant for a range of $\sim$ 0.3 eV around the Fermi level. Consequently, it is expected that this phase can be stable for some variation of both hole and electron doping, which may provide a way to stabilize it experimentally \cite{Feng2024}. For instance, raising the Te to Ta electron transfer by increasing the number of layers could be  a possibility. We remind that it has been proposed that this phase may coexist with the (3$\times$3) CDW in few-layers TaTe$_2$~\cite{Hwang2022}.    

The previous bonding scheme based on the occurrence of radial and peripheral Ta-Ta bonding also applies to the ($\sqrt 7\times \sqrt 7$) phase. The optimized structure contains 7-atom clusters with Ta$_\mathrm{I}$-Ta$_\mathrm{II}$ and Ta$_\mathrm{II}$-Ta$_\mathrm{II}$ distances of 3.27 \AA (practically identical to those in the ($\sqrt 13\times \sqrt 13$) phase) and Ta-Ta inter-cluster contacts of 3.64  and 4.25 \AA. The absence of the outer capping Ta atoms somewhat decreases the stabilization of the peripheral Ta-Ta bonding. However, the total Ta-Ta overlap population/f.u., i.e. the average Ta-Ta overlap population, is even stronger (13.6\%) than in the ($\sqrt {13}\times \sqrt {13}$) because all Ta atoms participate in the stronger radial and peripheral bonding. Why then the ($\sqrt 7\times \sqrt 7$) phase is 65 meV/f.u. less stable (Table~\ref{tab:energies})? In the absence of Ta$_\mathrm{III}$ atoms separating the 7 atom clusters, the Te atoms bridging two adjacent clusters are in a highly strained geometry that strongly destabilizes the Ta-Te network. For instance, the average Ta-Te overlap population/f.u. in ($\sqrt {13}\times \sqrt {13}$) is only 3.8\% weaker than in the very relaxed 1$T$ phase but 8.2\% weaker in the ($\sqrt 7\times \sqrt 7$) phase. Such uncomfortable situation in the inter-cluster region strongly decreases the phase stability despite the presence of quite stable 7-atom clusters. We conclude that such clusters are not likely to be found as ($\sqrt 7\times \sqrt 7$) CDW but as a part of more complex phases where they are separated by either isolated Ta atoms or other types of clusters (see next section) which will confer more freedom to the structure and thus reduce the strain due to clustering. The nature of the levels near the Fermi level is, however, very similar to those of the ($\sqrt {13}\times \sqrt {13}$) phase corresponding to a 2D metal (see the band structure in Fig. S4 and the FS in Fig.~\ref{fig:stars_FS}a).

The situation is very different for the higher members of the series. For instance, in  ($\sqrt {19}\times \sqrt {19}$)$_\mathrm{A}$ (see Fig.~\ref{fig:struct_2}c) there is not a clear separation between the inner and outer parts of the 19 atom Ta clusters. Although the Ta-Ta distances in the inner 7-atoms core are a bit shorter (3.33-3.35 \AA), they are longer than in the ($\sqrt 7\times \sqrt 7$) and ($\sqrt {13}\times \sqrt {13}$) phases. The Ta-Ta bonds in the outer ribbon of the cluster are in the range 3.36-3.52 \AA. This phase should be considered as a kind of local and somewhat irregular contraction of the Ta lattice of 1$T$-TaTe$_2$ combining a quite comfortable Ta-Te network but with less extensive Ta-Ta bonding. Thus, it is not surprising that this phase is calculated to be (see Table~\ref{tab:energies}) 47 meV/f.u. less stable than ($\sqrt {13}\times \sqrt {13}$) but 19 meV/f.u. more stable than ($\sqrt 7\times \sqrt 7$). The structural difference with the lower $m$ ($\sqrt m\times \sqrt m$) phases is clear when looking at the band structure (see Fig. S5a where only one band crosses the Fermi level) and even more clearly at the FS (Fig.~\ref{fig:stars_FS}c), which clearly reminds that of the non distorted 1$T$ phase, with three intersecting open but warped portions. Moreover, the Ta-Ta overlap population increases for a range of 0.25 eV just above the Fermi level so that electron-doping may be a way to stabilize this phase (see Fig. S5b).

\begin{figure*}[t]
    \centering   
    \includegraphics[scale=1]{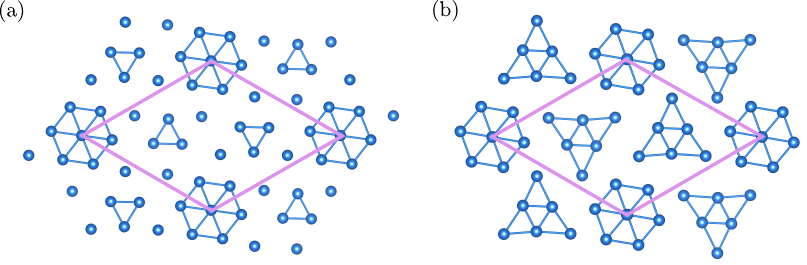}
    \caption{Two different ($\sqrt {19}\times \sqrt {19}$) single-layer TaTe$_2$ structures optimized in this work: ($\sqrt {19}\times \sqrt {19}$)$_\mathrm{B}$ (a) and ($\sqrt {19}\times \sqrt {19}$)$_\mathrm{C}$ (b). Only the Ta atoms are shown for simplicity.}
    \label{fig:struct_4}
\end{figure*}

\subsubsection{Structures with \texorpdfstring{$\sqrt m\times \sqrt m$} ~ units and spacers}\label{sec:stars and spacers}

In addition to ($\sqrt {19}\times \sqrt {19}$)$_\mathrm{A}$ we have found two further ($\sqrt {19}\times \sqrt {19}$)  phases shown in Fig.~\ref{fig:struct_4}. The first, ($\sqrt {19}\times \sqrt {19}$)$_\mathrm{B}$, is the metastable TaTe$_2$ phase recently stabilized in single- and few-layers TaTe$_2$  \cite{Hwang2022}. The structure of the 7-atom cluster (Ta$_\mathrm{I}$-Ta$_\mathrm{II}$ and Ta$_\mathrm{II}$-Ta$_\mathrm{II}$ distances of 3.23 \AA) is very similar to that in the ($\sqrt 7\times \sqrt 7$) phase (Ta$_\mathrm{I}$-Ta$_\mathrm{II}$ and Ta$_\mathrm{II}$-Ta$_\mathrm{II}$ distances of 3.27\AA). However, note that the Ta-Ta distances are now somewhat shorter, clearly showing that the strain is smaller because of the occurrence of Ta spacers and the metal-metal bonding within the 7-atom cluster can be optimized. Three of the twelve atoms in-between clusters are far apart from the 7-atoms clusters and form triangles with short Ta-Ta bonds of 3.0 \AA. The triangles and 7-atom clusters are separated by Ta spacers so that globally the strain is not very significant (only some strain around the triangular clustering occurs). 

It is very instructive to analyze the different contributions to the Ta-Ta bonding. As a matter of fact the Ta-Ta overlap population/f.u. is almost exactly that for the ($\sqrt 7\times \sqrt 7$) phase (the difference is smaller than 1~\%). The 7-atom clusters contribute 36.0~\% to this quantity, the triangles 27.1~\% and the inter-cluster contacts, 36.7~\%. Thus, the six Ta atoms in between clusters still contribute non negligibly because even if the associated Ta-Ta bonds are longer than those within the clusters they are shorter than in the 1$T$ phase. Note that the Ta-Ta bonding within the triangular units is provided by electron deficient 3c-2e bonds essentially similar but with a different topology (triangular $vs$ linear) to those discussed for the  (3$\times$1) phase. This is why even if the Ta-Ta distances are the shortest in the structure they are still longer than for typical localized 2c-2e bonds ($\sim$ 2.92 \AA)~\cite{Pyykko2009}. The contribution per Ta-Ta contact of the triangular units is in fact larger than that of the 7-atom units. In short, the ($\sqrt {19}\times \sqrt {19}$)$_\mathrm{B}$ structure is an efficient  way to build a relatively non-strained structure containing 7-atom clusters. By using a larger number of Ta atoms outside these 7-atom clusters, new and smaller cluster units can occur compensating the loss of radial and peripheral cluster bonding. The presence of Ta spacers insures that the strain due to the clustering is kept reasonably low. The final outcome is that this phase is 32~meV/f.u. more stable than the ($\sqrt {7}\times \sqrt {7}$) and although is still 50~meV/f.u. less stable than the (3$\times$3) phase, this quantity is reasonable low so that apparently it has been possible to stabilize it. 

\begin{figure*}
    \centering   
    \includegraphics[scale=0.9]{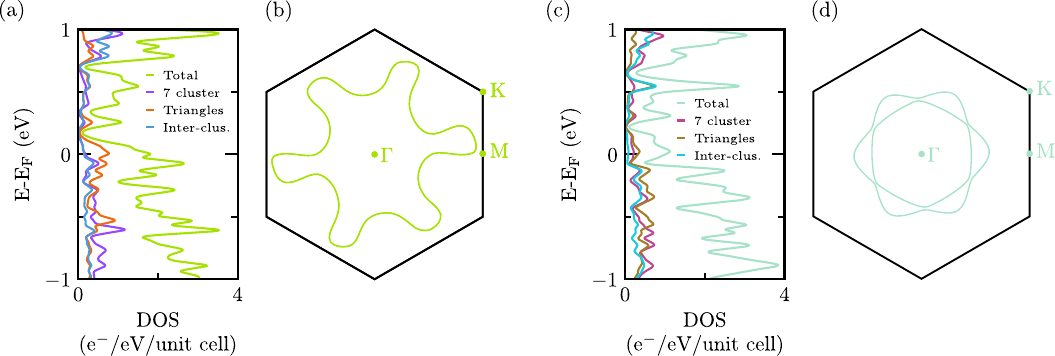}
    \caption{Calculated DOS and Fermi surface for the two ($\sqrt {19}\times \sqrt {19}$) phases of Fig.~\ref{fig:struct_4}. The results for ($\sqrt {19}\times \sqrt {19}$)$_\mathrm{B}$ are shown in (a)-(b) and those for ($\sqrt {19}\times \sqrt {19}$)$_\mathrm{C}$ in (c)-(d). The DOS plots also show the contribution of the 7-atom cluster, the two triangles (the two inner triangles of the 6-atom clusters for ($\sqrt {19}\times \sqrt {19}$)$_\mathrm{C}$) and the six inter-cluster Ta atoms (the six outer atoms of the 6-atom clusters for ($\sqrt {19}\times \sqrt {19}$)$_\mathrm{C}$). }
    \label{fig:B_vs_C}
\end{figure*}

The results concerning the ($\sqrt 13\times \sqrt 13$) and ($\sqrt 7\times \sqrt 7$) phases suggested us that there could be an even more effective way to stabilize this kind of phase with 7-atom clusters and triangles. The Ta$_\mathrm{III}$ atoms in the ($\sqrt 13\times \sqrt 13$) phase act as outer members of the 13-atom clusters: they provide some stabilization to the cluster levels just acting as electron accepting groups and this has the consequence of leading to noticeably larger Ta-Ta distances. Because of these lesser bonded Ta atoms, the lattice strain is strongly relieved and the phase is stabilized. Thus, we considered that the Ta spacers in the previous structure could be used in a more efficient way by capping either the 7-atom or the triangular clusters. Therefore, we looked for the two types of phases and finally could locate the ($\sqrt {19}\times \sqrt {19}$)$_\mathrm{C}$ phase (Fig.~\ref{fig:struct_4}b) where the triangular units are capped. Attempts to find phases with the 7-atom clusters capped led to the ($\sqrt {19}\times \sqrt {19}$)$_\mathrm{B}$ phase. The new phase is 15 meV/f.u. more stable than ($\sqrt {19}\times \sqrt {19}$)$_\mathrm{B}$ (Table \ref{tab:energies}) and is the more stable ($\sqrt 19\times \sqrt 19$) phase we have been able to locate.

The structure of the 7-atom cluster in this phase is practically the same as in ($\sqrt {19}\times \sqrt {19}$)$_\mathrm{B}$ (Ta$_\mathrm{I}$-Ta$_\mathrm{II}$ and Ta$_\mathrm{II}$-Ta$_\mathrm{II}$ distances of 3.24 \AA~$vs$ 3.23 \AA) and Ta-Ta distance of the inner triangle of the 6-atom cluster is somewhat shorter (2.94 $vs$ 3.0 \AA) because of the stabilization afforded by the capping. The two Ta-Ta distances between the capping Ta and the inner triangle are longer, 3.27 and 3.43 \AA, as they were in the 13-atom cluster of Fig.~\ref{fig:struct_2}. The Ta-Ta overlap population/f.u. is 3 \% larger than in ($\sqrt {19}\times \sqrt {19}$)$_\mathrm{B}$, as expected from the larger stability. Analysis of their different components show that the 7-atom cluster and triangles contribute around one-third (37.6 \%) and the inter-clusters contributes around two-thirds (62.4 \%). The individual contributions of the 7-atom cluster and the triangles are -17.2 \% and +54.8 \%, respectively. Consequently, the use of the inter-cluster Ta atoms capping the triangular units in the ($\sqrt {19}\times \sqrt {19}$)$_\mathrm{C}$ phase slightly disfavors the 7-atom cluster but considerably sabilizes both the triangles and the inter-cluster components. The capping has three consequences. First, stabilizes the 3c-2e bonds since the capping Ta atoms act as electron acceptors, as it happened in the ($\sqrt 13\times \sqrt 13$) phase (see Fig.~\ref{fig:star_scheme}b). Second, because of this electron accepting ability the Ta-Ta bonding is noticeably delocalized towards these capping atoms resulting in a more favorable interaction of the triangle and outer Ta atoms. Because of such delocalization, and the associated relatively short Ta-Ta bonds, the representation with 6-atom cluster (Fig.~\ref{fig:struct_4}b) provides a good representation of this phase. Third, because of the capping the local strain generated by the formation of the triangle clusters is reduced. The final outcome is that this phase is substantially more stable, only 23 meV/f.u. less stable than the ubiquitous (3$\times$1) phase. In view of the present results it remains as an open question if the previous metastable structure is really the ($\sqrt {19}\times \sqrt {19}$)$_\mathrm{B}$ phase or the ($\sqrt {19}\times \sqrt {19}$)$_\mathrm{C}$ one. Note that the different structural use of the Ta spacers is not simply a minor structural aspect but has a non negligible effect on the electronic structure. For instance, the calculated FS and the DOS for the two phases (Fig.~\ref{fig:B_vs_C}) clearly denote a quite different set of interactions around the Fermi level. The band structures of the two phases (Figs. S6 and S7 in SM) also exhibit significant differences around the Fermi level. Note that for both ($\sqrt {19}\times \sqrt {19}$)$_\mathrm{A}$ (Fig. S4) and ($\sqrt {19}\times \sqrt {19}$)$_\mathrm{B}$ (Fig. S6 in SM) quite flat bands cross the Fermi level so that it can not be discarded that Mott type effects compete with the CDW formation as for the ($\sqrt 13\times \sqrt 13$) 1$T$-TaX$_2$ (X= S, Se) phases. However, this is not the case for the more stable ($\sqrt {19}\times \sqrt {19}$)$_\mathrm{C}$ (Fig. S7 in SM). The reason lies on the different use of the inter-cluster Ta atoms around the Fermi level. Note in Fig.~\ref{fig:B_vs_C}a that the states of ($\sqrt {19}\times \sqrt {19}$)$_\mathrm{B}$ at the Fermi level are mostly on the triangles with a substantially smaller participation of the inter-cluster atoms so that these states are quite ``localized'' in the triangles and the dispersion of the bands is weak (Fig. S6 in SM). In contrast, for ($\sqrt {19}\times \sqrt {19}$)$_\mathrm{C}$ the states at the Fermi level are more evenly distributed among the different components denoting a larger delocalization, i.e. larger band dispersion (Fig. S7 in SM) and thus smaller values of the density of states.    

\subsubsection{4\texorpdfstring{$\times$}~1 and 4\texorpdfstring{$\times$}~4 structures}\label{sec:4x1_4x4}

As mentioned above the phonon dispersion for 1$T$ TaTe$_2$ (Fig.~\ref{fig:ep}) suggests the occurrence of some (4$\times$1) or (4$\times$4) instability. In fact, a very recent theoretical study \cite{Wang2024} proposed that a (4$\times$4) phase is the second lowest state for single-layer TaTe$_2$, just below the (3$\times$3) phase, and under hole doping it becomes the more stable phase. A (4$\times$4) phase was also theoretically predicted as the more stable phase for single-layer NbTe$_2$ \cite{Zhang2022} and experimental results confirmed the occurrence of both the (4$\times$1) and (4$\times$4) phases \cite{Bai2023}. More recently, a (4$\times$1) phase has been mentioned for single-layer TaTe$_2$ \cite{Feng2024}. Hence, we felt necessary to consider also this type of structures. 

\begin{figure*}[t]
    \centering   
    \includegraphics[scale=1]{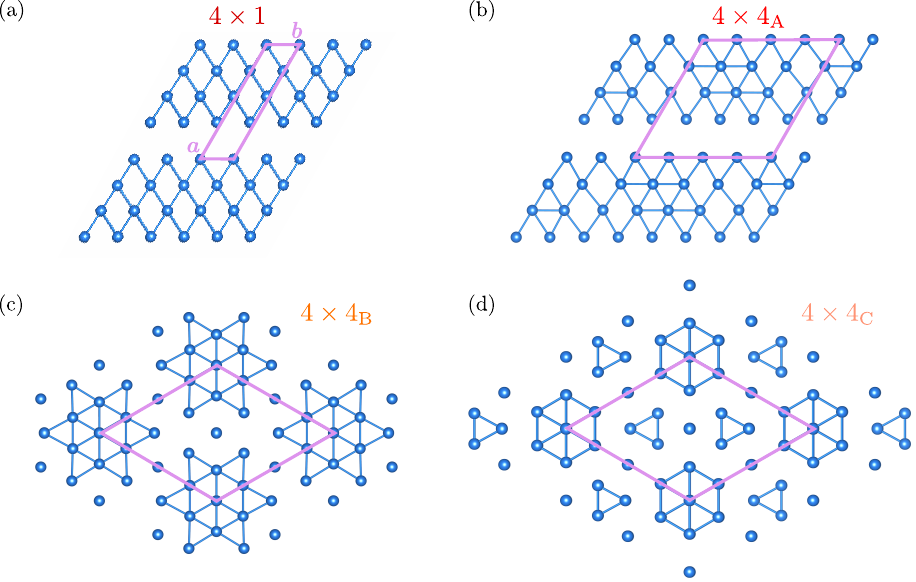}
    \caption{(4$\times$1) (a) and three different (4$\times$4) (b)-(d) single-layer TaTe$_2$ structures optimized in this work. Only the Ta atoms are shown for simplicity. }
    \label{fig:struct_5}
\end{figure*}

\begin{figure*}[t]
    \centering
    \includegraphics[scale=1]{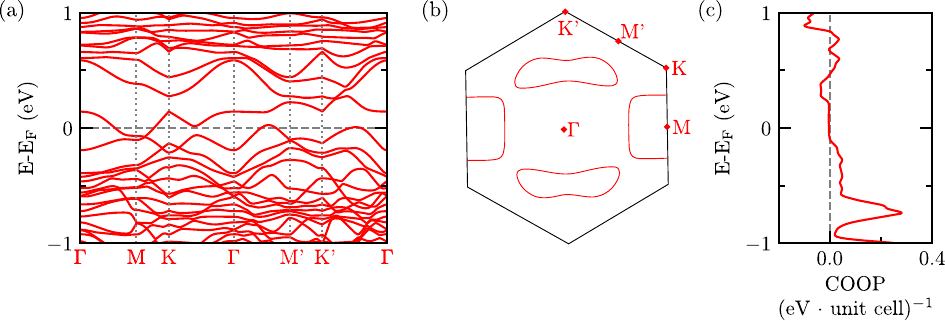}
    \caption{Band structure (a), Fermi surface (b) and COOP curve for the Ta-Ta bonds of the triple zigzag chain (c) of the (4$\times$4)$_\mathrm{A}$-TaTe$_2$ phase.  $\Gamma$, M, K, M' and K' refer respectively to the (0, 0), ($a$*/2, 0), ($a$*/3, $b$*/3), (0, $b$*/2) and (-$a$*/3, 2$b$*/3) points of the BZ.}
    \label{fig:4x4_A}
\end{figure*}

We have been able to locate one (4$\times$1) phase and three different (4$\times$4) phases (see Fig.~\ref{fig:struct_5}). The (4$\times$1) and (4$\times$4)$_\mathrm{A}$ phases are simply a generalization of the (3$\times$1) and (3$\times$3) phases discussed previously. The (4$\times$1) and (4$\times$4)$_\mathrm{A}$ phases can be described as resulting from a tetramerization along two and three directions, respectively, of the hexagonal lattice of the 1$T$ phase. The Ta-Ta distances in the chain of the (4$\times$1) phase (3.35-3.31-3.35 \AA) are longer than those in the chain of the (3$\times$1) phase (3.23-3.23\AA). This is also the case for the (4$\times$4)$_\mathrm{A}$ phase where the corresponding distances are 0.1-0.15 \AA~longer than in the (3$\times$3) phase. The (4$\times$1) phase is a bit less stable (8 meV/f.u.) than the (3$\times$1) phase (Table~\ref{tab:energies}) and is calculated to be metallic (see the band structure and FS in Fig. S8 of SM) as it is the (3$\times$1) phase. The (4$\times$4)$_\mathrm{A}$ phase is found to have practically the same stability as the (3$\times$3) phase (Table \ref{tab:energies}) in agreement with a previous work \cite{Wang2024}. The calculated band structure and FS are reported in Figs.~\ref{fig:4x4_A}a and b. The FS is reminiscent of that of the (3$\times$3) phase and can be considered as resulting from the hybridization of two components: a perpendicular one to the chain direction ($b$) and a closed one centered at $\Gamma$.

The bonding in this phase can be understood by a simple extension of the argument developed for the double zigzag chains of the (3$\times$1) and (3$\times$3) phases (Fig.~\ref{fig:3c_2e}). In the present cases there are triple zigzag chains so that the repeat unit of the chain contains 4 Ta atoms. Consequently, from every $t_{2g}$ levels four linear combinations can be built, two are bonding and two antibonding. The lowest bonding combination is an in-phase combination of the $t_{2g}$ orbital of the four Ta atoms (as it is the lowest combination in Fig.~\ref{fig:3c_2e}). When the system has two electrons to fill this orbital, the Ta-Ta distances will shorten and we can talk of a four-center two-electron bond (4c-2e). Even if there is no electron transfer, the Ta atoms are in a formal $d^1$ configuration so that there are four electrons to fill the lowest bonding combination built from two different orbitals, exactly as we discussed for the chains in the (3$\times$1) phase. Of course, the stabilisation afforded by a 4c-2e bond is smaller than for 3c-2e bond. This is why the bond lengths of the (4$\times$1) and (4$\times$4) phases are longer than for the (3$\times$1) and (3$\times$3) analogs. Again, there is one $t_{2g}$ per Ta atom that has not been used for bonding and the central atoms of the triple chain can use them to create additional bonding along the chain (i.e. the direction of the remaining $t_{2g}$ orbital) leading to the tetramerization along the chains of (4$\times$4)$_\mathrm{A}$ (Fig.~\ref{fig:struct_5}b).

Analysis of the overlap populations simply confirm this qualitative analysis. There is a large increase of the Ta-Ta total overlap population/f.u. of 30.4 \% for (4$\times$1) and 31.2 \% for (4$\times$4)$_\mathrm{A}$ with respect to the 1$T$ phase and such increase is smaller by $\sim$ 10 \% than that found for the trimerized phases. Since we are dealing with less stabilizing 4c-2e bonds, the shortening of the Ta-Ta distances in the chain are smaller and the strain generated is also smaller. For instance, the destabilization of the Ta-Te network judged from the Ta-Te overlap population/f.u. in the (4$\times$4)$_\mathrm{A}$ phase is 23.6 \% smaller than for the (3$\times$3) one. Thus, the relative stability of the trimerized vs tetramerized phases relies on a subtle deal between the stabilization afforded by the metal-metal bonding and the strain it generates. Shown in Fig.~\ref{fig:4x4_A}c is a COOP curve for the Ta-Ta bonds of the triple zigzag chains of (4$\times$4)$_\mathrm{A}$. The Ta-Ta bonding states are found in the lower part of the curve. What is interesting to remark is that there is a long range of energies of $\sim$ 1 eV around the Fermi level were the Ta-Ta overlap population is very small. This observation suggests that the metal-metal bonding remains almost constant in that energy region and, consequently, that the stability of the (4$\times$4)$_\mathrm{A}$ phase is not very sensitive to electron or hole doping. Since the repeat unit of the chain contains four Ta atoms, the system has the four electrons needed for the 4c-2e bonds without any doping. In contrast, the chain of the (3$\times$3) phase is more stable but needs the electron transfer from the Te bands to obtain the four electrons needed for the 3c-2e bonds. Therefore, hole doping or decreasing the Te to Ta electron transfer should have a stronger destabilizing effect on the (3$\times$3) phase. This provides a simple rationale for the previous suggestion that the (4$\times$4)$_\mathrm{A}$ phase becomes the most stable one under hole doping \cite{Wang2024}. It may also provide a simple explanation for the puzzling observation of the prevalence of trimerized (3$\times$1) and (3$\times$3) phases in both single-layer and bulk despite the fact that we (Table~\ref{tab:energies}) and others \cite{Wang2024} find the (4$\times$4)$_\mathrm{A}$ phase to be almost as stable as the (3$\times$3) one. Most likely the DFT calculations slightly underestimate the real Te to Ta electron transfer so that we are really in the region of coexistence of the two phases. In fact, as we mentioned  previously the calculated Te to Ta electron transfer is around 0.1 electrons which is a bit off to ensure the dominance of 2c-3e over 4c-2e bonds.     

The second most stable structure of this type is (4$\times$4)$_\mathrm{B}$ (Fig.~\ref{fig:struct_5}c) containing star-of-David clusters as well as isolated Ta atoms. This phase is only 8 meV/f.u. less stable than the ($\sqrt 13\times \sqrt 13$) one which matches the fact that the total overlap population/f.u. is only 3 \% smaller. (4$\times$4)$_\mathrm{B}$ is thus a quite stable phase being only 25 meV/f.u. less stable than the (3$\times$3) phase (Table~\ref{tab:energies}). As suggested for the ($\sqrt 13\times \sqrt 13$) phase \cite{Hwang2022} it is possible that this structure could be stabilized in few-layers TaTe$_2$. 

The structure of the 13-atom clusters is similar to that in the ($\sqrt 13\times \sqrt 13$) phase. The Ta-Ta distances in the inner 7-atom cluster are $\sim$0.02 \AA~longer whereas the outer Ta-Ta bonds are shorter ($\sim$0.04 \AA) and symmetric in the present case. Globally, the 13-atom clusters provide an overlap population/f.u. that is  20\% smaller than in the ($\sqrt 13\times \sqrt 13$) phase. However, the inter-cluster interactions almost compensate such decrease. Whereas in the  ($\sqrt 13\times \sqrt 13$) phase the inter-cluster contribution is very small, in the present case the isolated atoms provide a good connection between clusters (four Ta...Ta contacts of 3.61 \AA~per isolated Ta atom) and provide an overlap population/f.u. which is 22.1\% of the total. Finally, the stability of the two phases is not that different.

\begin{figure}[t]
    \centering   
    \includegraphics[scale=1]{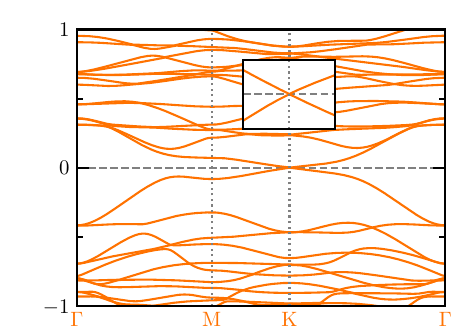}
    \caption{Band structure for the (4$\times$4)$_\mathrm{B}$-TaTe$_2$ phase where $\Gamma$= (0, 0), M= (1/2, 0) and K= (1/3, 1/3) in units of the reciprocal lattice vectors. A zoom-in around the Dirac cone from -0.03 to 0.03 eV is shown in the inset.}
    \label{fig:Dirac}
\end{figure}

The presence of the isolated Ta atoms in between the 13-atom clusters is very important. Although they do not significantly modify the stability of the phase they have a remarkable effect on the electronic structure. Since there are three isolated atoms, one per equivalent direction of the hexagonal lattice, they enforce a very symmetric structure. For instance, the three planes perpendicular to the single-layer running in the direction of the three main axes of the inner 7-atom cluster are kept whereas it is not the case for ($\sqrt 13\times \sqrt 13$) (compare Figs.~\ref{fig:struct_5}c and \ref{fig:struct_2}b). As discussed above, the connection between the clusters through the isolated atoms is quite sizeable. These facts completely change the nature of the electronic interactions and lead to a very different band structure (Fig.~\ref{fig:Dirac}) containing a Dirac cone just at the Fermi level as a result of the symmetry properties of the lattice. Although the crossing will not survive the introduction of spin-orbit coupling because the plane of the Ta atoms is not a symmetry plane (Ta in octahedral coordination), the occurrence of such Dirac cone calls for detailed attention towards this phase given its good stability. Although it is out of the scope of this work, a study of the possible topological properties, stability, and dependence of the nature of both transition metal and chalcogen atoms among group V TMDC is now under way \cite{Dirac}.  

Finally, we located the very symmetric (4$\times$4)$_\mathrm{C}$ phase shown in Fig.~\ref{fig:struct_5}d. In this structure, the six outer Ta atoms of the 13-atom cluster of the previous phase have been used to introduce two additional triangles in between clusters. The phase is also closely related to the ($\sqrt {19}\times \sqrt {19}$)$_\mathrm{B}$ phase (Fig.~\ref{fig:struct_4}a) which contains three additional isolated Ta atoms. Surprisingly, this phase is one of the less stable among those we have located, being only 17 meV/f.u. more stable than the 1$T$ phase (Table~\ref{tab:energies}). 

This apparently surprising result is however easy to understand. Both the 7-atom cluster and the triangles have relatively long Ta-Ta distances (for instance the 7-atom clusters are $\sim$0.20 \AA~longer than in ($\sqrt {19}\times \sqrt {19}$)$_\mathrm{B}$ and thus have a considerably diminished contribution to the Ta-Ta overlap population/f.u. These longer distances are partially due to the symmetry constraints imposed by the lattice with two different types of clusters. In addition, the contribution of the isolated Ta atoms is also small because they do not really act as capping atoms. Remind that capping atoms provide a very sizeable stabilization to the system by lowering the energy of the cluster levels through the delocalization toward the capping atoms (see Fig.~\ref{fig:star_scheme}b) and by lowering the strain. Here the isolated Ta atoms between the triangles are not very effective at capping because they would be shared by two triangles. The possibility to really cap one of the two triangles is not allowed by the symmetry of the lattice. In addition, the 7-atom cluster can not be capped because the appropriate Ta atoms are used to build the triangles. Note that in the ($\sqrt {19}\times \sqrt {19}$)$_\mathrm{B}$ phase the 7-atom clusters are surrounded by a set of six isolated Ta atoms in the correct positions for capping. Although the distances are a bit longer than in usual 13-atom clusters they are shorter than in the 1$T$ structure and they have a partial capping role. We thus believe that this phase is too high in energy to be experimentally prepared, although it provides a clear illustration of the key role of Ta atom spacers in stabilizing a given structure.

\subsubsection{Further structures}\label{sec:further}

\begin{figure}[t]
    \centering   
    \includegraphics[scale=1]{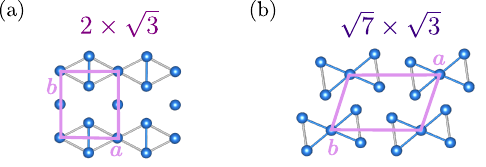}
    \caption{(2$\times \sqrt {3}$) and ($\sqrt {7} \times \sqrt {3}$) single-layer TaTe$_2$ structures studied in this work. Only the Ta atoms are shown for simplicity. The grey bonds become longer than 3.45 \AA~in the optimized structures (see text).}
    \label{fig:struct_7}
\end{figure}

Finally, we consider two different structures where the Ta atoms occur as fused triangular units. The (2$\times \sqrt {3}$) structure (Fig.~\ref{fig:struct_7}a) contains chains of edge- and vertex-shared Ta triangular units whereas the ($\sqrt {7} \times \sqrt {3}$) one (Fig.~\ref{fig:struct_7}b) contains vertex-shared double triangular units. A previous study \cite{Si2020} has shown that for single-layer VSe$_2$ these phases are comparable in stability to that built from the star-of-David clusters of Fig.~\ref{fig:struct_5}c, i.e. (4$\times$4)$_\mathrm{B}$. In fact, except under compressive strain, the ($\sqrt {7} \times \sqrt {3}$) structure is the more stable phase \cite{Si2020}. Since the (4$\times$4)$_\mathrm{B}$ is only 25 meV/f.u. less stable than the (3$\times$3) one we felt interesting to look for the possible (2$\times \sqrt {3}$) and ($\sqrt {7} \times \sqrt {3}$) structures of single-layer TaTe$_2$.

We were unable to find any of the two phases. Our optimised structures led to phases where the bonds in grey in Fig.~\ref{fig:struct_7} are long. For instance, in the (2$\times \sqrt {3}$) structure the blue bond is 3.25 \AA~whereas the grey bonds are 3.52 \AA~and the contacts with the isolated Ta atoms are 3.72 \AA. Hence, the Ta network is simply a set of quasi-independent dimers and isolated Ta atoms. In the case of the ($\sqrt {7} \times \sqrt {3}$) phase the blue bonds are 3.40-3.42 \AA~and the longer grey bonds, 3.54 \AA. The contacts between adjacent triangular units are 3.49 \AA~along $b$ and 3.57\AA~along $a$. Thus, the Ta network of this phase can be considered to contain 5-atom fragments of the chains in the very stable (3$\times$1) structure somewhat interconnected along the $b$ direction. 

As expected from the non occurrence of an extensive set of metal-metal bonds the ($\sqrt {7}\times\sqrt {3}$) and (2$\times \sqrt {3}$) phases rank among the less stable phases considered. In fact, the  ($\sqrt {7} \times \sqrt {3}$) phase is only 28 meV/f.u, more stable than the 1$T$ phase and the (2$\times \sqrt {3}$) phase is even 10 meV/f.u. less stable than the 1$T$ phase. The Ta-Ta overlap population/f.u. is only 9.3~\% [(2$\times \sqrt {3}$)] and 9.5~\% [($\sqrt {7} \times \sqrt {3}$)] larger than in the 1$T$ phase, i.e. between 5 and 6 times smaller than in the more stable phases and this weak stabilization afforded by the metal-metal bonding is almost compensated [($\sqrt {7} \times \sqrt {3}$)] or even superseded [(2$\times \sqrt {3}$)] by the local strains generated around the meta-metal bonds.

These results provide a vivid illustration of the richness of the structural landscape of group V TMDC single-layers. Whereas the ($\sqrt {7} \times \sqrt {3}$) phase is the more stable phase for VSe$_2$ \cite{Si2020} and only under compressive strain the (4$\times$4)$_\mathrm{B}$ phase can become more stable, for TaTe$_2$ single-layers the ($\sqrt {7} \times \sqrt {3}$) phase is 78 meV/f.u. less stable than the (4$\times$4)$_B$!

\section{Concluding remarks}

The CDWs in single-layer 1$T$ MTe$_2$ (M= Nb, Ta) have recently raised a large attention because of the contrasting behavior with those in the sulfide and selenide analogues. Both NbTe$_2$ and TaTe$_2$ exhibit several modulations with complex unit cells and no clear rational for this diversity and the factors favoring the different modulations has emerged. The Ta atoms in these phases have formally a $d^1$ configuration and there are not enough electrons to explain the structure of the different CDW containing networks of Ta atoms exhibiting double or triple zigzag chains, star-of-David clusters, etc. In this work fourteen different single-layer phases of 1$T$-TaTe$_2$ have been studied by means of DFT calculations. Using concepts developed to explain the chemical bonding in molecular electron deficient systems and the observation that there is some Te to Ta electron transfer it has been possible to rationalize the origin of the different structures. 

The main picture emerging from this study is the following.  
Because of the very diffuse orbitals of the Te atoms there is a substantial overlap between the top of the tellurium-based valence bands and the bottom of the transition metal-based conduction bands. Such overlap is thus associated with a Te to Ta electron transfer. Since the main factor contributing to the stabilization of the lattice under the CDW structural modulation is the additional metal-metal bonding, the Te to Ta electron transfer has a strong control of the nature of the adopted CDW.
The semimetallic band overlap and thus the Te to Ta electron transfer is easily tunable with small structural changes. For instance, slightly decreasing the $a$ constant or progressively introducing inter-layer interactions as in bilayers, trilayers or bulk increases the electron transfer. Therefore, the group V ditellurides have a subtle mechanism to vary the occupation of the transition metal based levels and thus induce different ways to stabilize the Ta network through the CDW modulation. This is the essential difference with the disulfide an diselenide analogs where, because of the smaller extension of the Se 4$p$ and S 3$p$ orbitals, the semimetallic overlap is very small or absent and does not influence the choice between possible CDW. Consequently, both the disulfide and diselenide adopt the ($\sqrt {13}\times \sqrt {13}$) CDW which is just one of the possibilities for the ditelluride. As discussed in detail along this work, there are other factors at work but the Te to Ta electron transfer seems to play the main role.

All the systems we have examined exhibit structural distortions in the optimized structure which are considerably stronger than those occurring for CDW with a weak-coupling FS nesting based mechanism. In addition, the analysis of the FS and Lindhard response of the 1$T$ pristine phase does not provide compelling evidence for a nesting-based mechanism of the instabilities. Altogether, these facts lead to the conclusion that the driving force for the observed CDWs must be related to phonon instabilities or electron-phonon $k$-dependence.   

 The more stable phases for the single-layer 1$T$-TaTe$_2$ are found to be the (3$\times$3) and (4$\times$4) CDWs, the former being marginally more stable. However, other modulations like (3$\times$1), another (4$\times$4), (4$\times$1) and ($\sqrt {13}\times \sqrt {13}$) occur within a small energy range of 20 meV/f.u. above the previous ones. This suggests the possibility to prepare and characterize them. Some of these, like the (4$\times$4) and (4$\times$1), have been suggested in experimental studies and our theoretical study could help their characterization.  We note that metastable ($\sqrt {19}\times \sqrt {19}$) phases have also been reported. The more extensively studied one has been proposed to have a structure different from the more stable of these phases that we have located. Thus, it is an open question if several ($\sqrt {19}\times \sqrt {19}$) CDWs with different structures can exist. This situation was previously found for 1$H$-NbSe$_2$ \cite{Guster2019}.  

 The present study shows that although the ($\sqrt {13}\times \sqrt {13}$) modulation is quite stable, the smaller member of the family, ($\sqrt {7}\times \sqrt {7}$), is considerably less stable and it is not expected to be experimentally detected. However, the clusters of this phase are found as part of the more complex CDWs like ($\sqrt {19}\times \sqrt {19}$) or (4$\times$4). According to our calculations these phases usually contain clusters of 7- or 13-atoms, triangles and separate Ta atoms. These Ta atoms have a very important role in stabilizing the different structures depending on their ability to act as capping the different clusters. 
 
 Although all the CDW phases that we have studied exhibit band structures typical of metallic materials we call attention toward one of them, the second more stable (4$\times$4) one, (4$\times$4)$_\mathrm{B}$. This system exhibits a Dirac cone at the Fermi level and thus anticipates a potentially interesting physical behavior. Other phases exhibit both very flat and dispersive bands at the Fermi level so that Mott effects could also develop and provide an interesting behavior.

The stability of the different CDWs examined in this work correlates directly with the bonding strength of the Ta network which can be easily evaluated through the overlap population. Consideration of the metal-metal bonding as well as the local strain generated provide a clear understanding of these otherwise complex materials. The COOP curves offer a pictorial way to locate the states associated with Ta-Ta bonding and can be of great help in understanding the stability of different phases in simple terms and how to modify it through hole or electron doping. For instance, the analysis of the Ta-Ta bonding in the almost equally stable (3$\times$3) and (4$\times$4) CDWs suggest that the (4$\times$4) one could be stabilized and become the more stable phase by hole doping or decreasing the Te to Ta electron transfer. The approach developed here blending first-principles calculations, tools like the COOP curves and concepts from the bonding theory of electron deficient systems provides a useful way to progress in our understanding of systems with several CDWs as the group V ditellurides.


\section{Computational details}

First-principles calculations were carried out using a numerical atomic orbitals approach to density functional theory (DFT)~\cite{kohsha1965,HohKoh1964}, which was developed for efficient calculations in large systems and implemented in the \textsc{Siesta} code.~\cite{ArtAng2008,SolArt2002} We have used the generalized gradient approximation (GGA) and, in particular, the functional of Perdew \emph{et al.}~\cite{pbesol}.
Only the valence electrons are considered in the calculation, with the core being replaced by norm-conserving scalar relativistic pseudopotentials~\cite{vansetten2018dojo,garcia2018psml} factorized in the Kleinman-Bylander form~\cite{klby82}. The non-linear core-valence exchange-correlation scheme~\cite{LFC82} was used for all elements. We have used a split-valence double-$\zeta $ basis set including polarization functions~\cite{arsan99}. The energy cutoff of the real space integration mesh was set to 500 Ry. 
To build the charge density and, from this, obtain the DFT total energy and atomic forces, the Brillouin zone (BZ) was sampled with the Monkhorst-Pack scheme~\cite{MonPac76} using optimized grids.
For the Fermi surface (FS), density of states (DOS) and Crystal Orbital Overlap Population (COOP) calculations, a finer grid was used.
The forces and energies do not change significantly with the finer grid, which confirms the appropriateness of the grid for the relaxation as well as for the energies comparison.
For each structure, we optimized the lattice vectors as well as the atomic positions with a threshold of 0.01 eV/\AA.
Most of the structures and Fermi surfaces were plotted using VESTA~\cite{vesta} and XCRYSDEN~\cite{xcrysden}, respectively.
To calculate the phonon band structure of the 1$T$ structure, we performed Density Functional Perturbation Theory (DFPT) calculations as implemented in Quantum Espresso \cite{QE-2009,QE-2017}. 
We use ultrasoft pseudopotentials \cite{US-pseudos,QE-pseudos}, an optimized value for the plane-wave cutoff energy of 140 Ry and a $(18\times18\times1)$ Monkhorst-Pack grid.
The phonon $q$-grid was set to $(9\times9\times1)$.

\section*{Acknowledgments}
IMDEA Nanociencia acknowledges support from the ‘Severo Ochoa’ Program for Centres of Excellence in R\&D (CEX2020-001039-S/AEI/10.13039/501100011033).
J.A.S.-G. acknowledges support from NOVMOMAT, project PID2022-142162NB-I00 funded by MICIU/AEI/10.13039/501100011033 and by FEDER, UE.
E.C. was supported by the Spanish MCIN/AEI/10.13039/501100011033 and AEI through Grant PID2022-139776NB-C61 and the Severo Ochoa FUNFUTURE MaTrans42 (CEX2023-0001263-S) Excellence Centre distinction, as well as by Generalitat de Catalunya (Grant No. 2021 SGR 01519).
We thankfully acknowledge RES resources provided by SCAYLE in Calendula to FI-2020-2-0040 and BSC in MareNostrum5 to FI-2024-2-0009.

\bibliography{tate2-bib}

\end{document}